\documentclass[%
 reprint,
 amsmath,amssymb,
 aps,
prb,
]{revtex4-2}

\usepackage{graphicx}% Include figure files
\usepackage{dcolumn}% Align table columns on decimal point
\usepackage{bm}% bold math
\usepackage{xcolor}
\begin{document}

%\preprint{APS/123-QED}

\title{Series Expansion of a Scalable Hermitian Excitonic Renormalization Method}

\author{Marco Bauer$^1$}
\email[]{marco.bauer@iwr.uni-heidelberg.de}
\author{Andreas Dreuw$^1$}
\email[]{dreuw@uni-heidelberg.de}
\author{Anthony D. Dutoi$^2$}
\email[]{adutoi@pacific.edu}
%\homepage[]{https://wwwagdreuw.iwr.uni-heidelberg.de/, http://copresearch.pacific.edu/adutoi}
\affiliation{$^1$Interdisciplinary Center for Scientific Computing, Ruprecht-Karls University, Im Neuenheimer Feld 205, 69120 Heidelberg, Germany}
\affiliation{$^2$Department of Chemistry, University of the Pacific, Stockton, CA 95204, USA}

\date{\today}% It is always \today, today,

\begin{abstract}
\noindent
Utilizing the sparsity of the electronic structure problem, fragmentation methods have been researched for decades with great success, pushing the limits of \textit{ab initio} quantum chemistry ever further.
Recently, this set of methods was expanded to include a fundamentally different approach called excitonic renormalization, providing promising initial results.
It builds a supersystem Hamiltonian in a second-quantized-like representation from transition-density tensors of isolated fragments, contracted with biorthogonalized molecular integrals.
This makes the method fully modular in terms of the quantum chemical methods applied to each fragment and enables massive truncation of the state-space required. 
Proof-of-principle tests have previously shown that an excitonically renormalized Hamiltonian can efficiently scale to hundreds of fragments, but the \textit{ad hoc} approach to building the Hamiltonian was not scalable to larger fragments.
On the other hand, initial tests of the originally proposed modular Hamiltonian build, presented here, have shown the accuracy to be poor on account of its non-Hermitian character.
In this study, we bridge the gap between these with an operator expansion that is shown to converge rapidly, tending towards a Hermitian Hamiltonian while retaining the modularity, yielding an accurate, scalable method.
The accuracy is tested here for a beryllium dimer.
At distances near equilibrium and longer, the zeroth-order method is comparable to CCSD(T), and the first-order method to FCI.
The second-order method agrees with FCI for distances well up the inner repulsive wall of the potential.
Deviations occurring at shorter bond distances are discussed along with approaches to scaling to larger fragments.
\end{abstract}


\maketitle

%\tableofcontents

\section{\label{sec:intro}Introduction}

Fragmentation schemes were explored in many different flavors throughout the history of quantum chemistry. Up to this date they are heavily researched, since they provide a great opportunity to expose the sparsity of the exact solution of the Hamiltonian. Often a classical interaction with the environment suffices already in describing the perturbation of the environment on the subsystem of interest. An abundance of methods has been developed for this purpose, incorporating the effects of the environment from just point-charges, up to high orders of multipoles.\cite{Tomasi.1994, Olsen.2010, Rahbar.2023, Jacob.2014, Knizia.2012, Inglesfield.1981}

Fragmentation methods, also accounting for exchange and correlation, provide a similar variety,\cite{Jeziorski.1994, Shen.2009, Parker.2013, Marti.2010, Kitaura.1999, Yang.1991} but most of them either perform well for niche applications only or are difficult to generalize. 
Interestingly most of them bear different bottlenecks, indicating that fragmentation schemes do not share a common bottleneck, which needs to be overcome, but we rather require an ansatz that acts as a sweet spot between the common bottlenecks of fragmentation schemes. Approaches like block correlated coupled cluster (BCCC)\cite{Shen.2009, Li.2004, Wang.2020} and active space decomposition (ASD)\cite{Parker.2013} methods require explicit orbital orthogonalization for all fragments, while symmetry-adapted perturbation-theory (SAPT)\cite{Jeziorski.1994} requires a tedious derivation for global anti-symmetry for more than two fragments, as well as high computational effort starting from consistent third order. Roughly two decades ago, also the density-matrix renormalization-group (DMRG)\cite{Marti.2010, Baiardi.2020} has been introduced to the field of quantum chemistry with great initial success, remaining one of the best performing approximations to full configuration interaction (FCI) up to this date. However, DMRG performs poorly in terms of dynamic correlation. Hence, it is mostly applied to active spaces, pushing their limits in terms of number of electrons and active orbitals by an order of magnitude.

Recently, another \textit{ab initio} fragmentation scheme has been proposed and tested in proof-of-principle work.\cite{Dutoi.2019, Liu.2019} It was shown that the Hamiltonian could be written without approximation in terms of operators that induce fluctuations between correlated states of constituent fragments. Importantly, it was shown that the necessary Hamiltonian matrix elements could be computed in a systematically improvable way, using only fragment-local information.
The quantities required to set up the matrix elements of the Hamiltonian boil down to  intrafragment density tensors for correlated states, bearing no restrictions on the method used to calculate them, as well as interfragment one- and two-electron integrals.
Note
 that the intrafragment densities are reusable while also strictly separating short- and long-range correlation. Therefore, it was asserted that truncation of the model spaces for the fragments should provide a path forward for efficient computation. The final Hamiltonian is therefore called an excitonically renormalized (XR) Hamiltonian, with reference to the local site basis of Frenkel–Davydov exciton theory.

Since renormalization is also the central aspect in DMRG, it is useful to point out a few important differences between XR and DMRG. The most important difference is the way the fragments are connected in terms of their interactions. DMRG maps the fragments into a one dimensional chain of fragments, while XR yields a fully connected network. 
As an immediate consequence, the tensor-product states in DMRG need to be solved for in an iterative procedure, solving the fragment chain forward and backward, until convergence is met. Building the XR Hamiltonian itself on the other hand does not require an iterative procedure. Furthermore, DMRG depends on how the fragments are ordered in the one dimensional chain, since each subsystem only interacts with the one to its ``left'' or ``right.''  If another important interaction 
connects distant parts of the abstract chain, 
it might fall below the threshold of the local density-matrix truncation, since the system and the environment block are both heavily truncated. XR on the other hand includes all pairwise interactions into the two body part of the Hamiltonian, all interactions between three fragments, which are not covered by pairwise interactions, into the three body part, \textit{etc}. Another important feature of XR is the fact that the densities can be naturally obtained from a method of choice,
 whereas DMRG is not flexible about the level of theory allowed for the fragments.

Since XR is formulated in a second-quantized-like representation, one can derive analogues to the methods used to handle the standard Hamiltonian in a straightforward manner. An analogue to coupled-cluster with single and double substitutions (CCSD) has already been derived within the fluctuation operator representation, called XR-CCSD.\cite{Liu.2019}
For an excitonic Hamiltonian with up to dimer interactions, XR-CCSD formally scales with the third power of the number of fragments and the fourth power of the number of states per fragment, not including the building of the Hamiltonian.  Accounting for the sparsity of the effective Hamiltonian, actual scalings are even better than the formal scalings, outperforming the scaling of sixth-order with the system size for standard CCSD by more than three orders.
Excited state methods can also be derived in the same straightforward manner, and, since a wavefunction is obtained in addition to an energy, in contrast to incremental schemes,\cite{Nolan:2010:Hierarchical} 
properties are also naturally accessible.
Note
 that conventional CCSD is known to reproduce dynamic correlation very well, so XR-CCSD
is expected to do so as well.
Indeed, a variant of XR-CCSD (XR$^\prime$-CCSD) was shown to reproduce the FCI results of a model system very well with only a few states per fragment (Be atoms), and scale easily to 100 fragments, showing that the previous assertion of truncating the state space for efficient and precise computations is valid.\cite{Liu.2019}

For purely pragmatic reasons, the Hamiltonian as derived in the first publications was not implemented as originally described, since, for the small Be test systems involved, it could easily be extracted from FCI calculations on dimers.
This additional \textit{ad hoc} approximation was given the designation XR$^\prime$.
It is not scalable, however, and it was therefore desired to build and test the excitonic Hamiltonian according to the original XR formulation.

In this article, we first show what happens when the Hamiltonian is implemented exactly as described in the first part of the prior work. Due to its non-Hermitian character, the results are actually worse than for the \textit{ad hoc} XR$^\prime$ approximation. However, we herein put the latter approximation on rigorous theoretical grounds, expressing it as the limit of a systematically improvable series with tractable low-order terms.
Initial numerical results show that it appears to converge quickly, giving rise to an accurate and scalable build of the XR Hamiltonian in a heavily truncated basis.

\section{\label{sec:theory}Theory}

\subsection{\label{sec:theory:background}Review of Excitonic Renormalization}

This article closely follows the notational conventions of previous publications,\cite{Dutoi.2019, Liu.2019} with deviations given in Appendix \ref{sec:notational_diffs}.
The complete Fock space of a supersystem (within a given one-electron representation) is spanned by a basis $\{|\Psi_I\rangle\}$, with states
\begin{eqnarray}
    |\Psi_I\rangle = |\psi_{i_1}\psi_{i_2}\cdots\psi_{i_N}\rangle
\end{eqnarray}
where $I=(i_1, i_2, \cdots i_N)$ is an ordered tuple that specifies the states of the $N$ constituent fragments. The set $\{|\psi_{i_m}\rangle\}$ spans the complete Fock space for fragment $m$.
These states are general and can be chosen to describe the strong electron correlations inside of the individual fragments. 
The placement of the single-fragment state labels inside of a single ket indicates that the electronic state of the supersystem is overall antisymmetric.
A central concern is the insistence that the states for each fragment are defined in isolation, independent of their context in any specific supersystem.
The orbitals used to describe them therefore overlap those of their neighbors, and the basis $\{|\Psi_I\rangle\}$ is not orthonormal.

Via the introduction of a biorthogonal complement basis $\{|\Psi^I\rangle\}$ ($\langle\Psi^I|\Psi_J\rangle = \delta_{IJ}$), a set of fluctuation operators $\{\hat{\tau}_{i_m}^{j_m}\}$ can be defined that induce fragment $m$ in a supersystem state to change from state $j_m$ to $i_m$, assuming it was in state $j_m$ to begin with (otherwise yielding zero).  This then leads to an exact rewriting of the \textit{ab initio} Hamiltonian
\begin{eqnarray}
    \label{eq:abinitio}
    \hat{\mathcal{H}}
    &=&
      \sum_{pq} t^p_q \hat{c}_p \hat{a}^q 
    + \sum_{\alpha pq} {}^{\alpha}u^p_q \hat{c}_p \hat{a}^q 
    + \sum_{pqrs} v^{pq}_{rs} \hat{c}_p \hat{c}_q \hat{a}^s \hat{a}^r
    \\ &=&
    \label{eq:excitonic}
      \sum_{\substack{\mathbf{m}~=\\ \{m\}}} \sum_{\substack{I=(i_m)\\J=(j_m)}}
      \langle\Psi^I|\hat{H}_{\mathbf{m}}|\Psi_J\rangle
      \hat{\tau}^{j_m}_{i_m}
    \\ &~&
    + \sum_{\substack{\mathbf{m}~=\\ \{m_1,m_2\}}} \sum_{\substack{I=(i_{m_1}, i_{m_2})\\J=(j_{m_1}, j_{m_2})}}
      \langle\Psi^I|\hat{H}_{\mathbf{m}}|\Psi_J\rangle
      \hat{\tau}^{j_{m_1}}_{i_{m_1}} \hat{\tau}^{j_{m_2}}_{i_{m_2}} \nonumber \\
    &~&
    + \sum_{\substack{\mathbf{m}~=\\ \{m_1,m_2,m_3\}}} \sum_{\substack{I=(i_{m_1}, i_{m_2}, i_{m_3})\\J=(j_{m_1}, j_{m_2}, j_{m_3})}} \nonumber \\
    &~& ~~~~~~~~~~~~~~~~~~~~~
      \langle\Psi^I|\hat{H}_{\mathbf{m}}|\Psi_J\rangle \hat{\tau}^{j_{m_1}}_{i_{m_1}} 
      \hat{\tau}^{j_{m_2}}_{i_{m_2}} \hat{\tau}^{j_{m_3}}_{i_{m_3}} \nonumber \\
    &~&
    + \sum_{\substack{\mathbf{m}~=\\ \{m_1,m_2,m_3,m_4\}}} \sum_{\substack{I=(i_{m_1}, i_{m_2}, i_{m_3}, i_{m_4})\\J=(j_{m_1}, j_{m_2}, j_{m_3}, j_{m_4})}} \nonumber \\
    &~& ~~~~~~~~~~~~~~~
      \langle\Psi^I|\hat{H}_{\mathbf{m}}|\Psi_J\rangle \hat{\tau}^{j_{m_1}}_{i_{m_1}} 
      \hat{\tau}^{j_{m_2}}_{i_{m_2}} \hat{\tau}^{j_{m_3}}_{i_{m_3}} \hat{\tau}^{j_{m_4}}_{i_{m_4}} \nonumber
\end{eqnarray}
Here, the molecular integrals have been denoted
\begin{eqnarray}
    t^p_q &=& \langle\chi^p|\hat{t}|\chi_q\rangle \\
    {}^{\alpha}u^p_q &=& \langle\chi^p|{}^{\alpha}\hat{u}|\chi_q\rangle \\
    v^{pq}_{rs} &=& \frac{1}{4}\langle\chi^p\chi^q|\hat{v}|\chi_r\chi_s\rangle
\end{eqnarray}
where $\hat{t}$ is the single-electron kinetic energy, ${}^{\alpha}\hat{u}$ accounts for attraction of electrons to the nucleus of atom $\alpha$, and $\hat{v}$ is the electron–electron repulsion operator.
Placing multiple orbitals within a single bra or ket denotes a Slater determinant, and the members of the set $\{|\chi_p\rangle\}$ are the original overlapping fragment-localized orbitals.
The field operators $\hat{c}_p$ and $\hat{c}^p$ create the orbital states $|\chi_p\rangle$ and $|\chi^p\rangle$, respectively, which are members of the biorthogonal bases for the one-electron space ($\langle\chi^p|\chi_q\rangle=\delta_{pq}$), and $\hat{a}_p$ and $\hat{a}^p$ are their respective Hermitian conjugates.
Creation operators from one basis only obey canonical anticommutation relationships with annihilation operators from the other.

Eq.~\ref{eq:excitonic} references a decomposition of the \textit{ab initio} resolution of the Hamiltonian
\begin{eqnarray}
    \label{eq:abinitiofragdecomp1}
    \hat{\mathcal{H}} = \hat{H}_1 + \hat{H}_2 + \hat{H}_3 + \hat{H}_4
\end{eqnarray}
where, for example
\begin{eqnarray}
    \label{eq:abinitiofragdecomp2}
    \hat{H}_2
    ~=
    \sum_{\substack{\mathbf{m}~=\\ \{m_1,m_2\}}} \hat{H}_{\mathbf{m}}
\end{eqnarray}
is a sum over all unique dimers of fragments.
The terms $\hat{H}_{\{m\}}$ that comprise $\hat{H}_1$ collect together the terms from the $\hat{t}$, $\hat{u}$, and $\hat{v}$ parts of eq.~\ref{eq:abinitio} that have all indices (nuclei included) pertaining fragment $m$, and $\hat{H}_{\{m_1,m_2\}}$ similarly collects all terms for the dimer of fragments $m_1$ and $m_2$, excluding those already in $\hat{H}_{\{m_1\}}$ and $\hat{H}_{\{m_2\}}$, and so on.
The two-electron nature of the \textit{ab initio} Hamiltonian causes the fragment-based resolution in eq.~\ref{eq:excitonic} to truncate rigorously with tetramer terms because moving two electrons touches maximally four fragments.

Eq.~\ref{eq:excitonic} provides an in-principle exact framework that reformulates the supersystem electronic-structure calculation in terms of correlated fluctuations of fragments, assuming the matrix elements that comprise the scalar coefficients of the expansion can be resolved.
It is enlightening to look at the working equations for just one such matrix element, for example $\langle\Psi^I|\hat{H}_{\{1,2\}}|\Psi_J\rangle$,
\begin{eqnarray}
    \label{eq:XRdimerelemtent}
    \langle\psi^{i_1} \psi^{i_2}|\hat{H}_{\{1,2\}}|\psi_{j_1} \psi_{j_2}\rangle
    &=&   \langle {}^{2}u^{1}_{1} \rangle \delta_2
        + \langle {}^{1}u^{2}_{2} \rangle \delta_1
        + \langle t^{1}_{2} \rangle + \langle t^{2}_{1} \rangle
        \nonumber \\
    &~& +~\langle {}^{1}u^{1}_{2} \rangle + \langle {}^{1}u^{2}_{1} \rangle
        + \langle {}^{2}u^{1}_{2} \rangle + \langle {}^{2}u^{2}_{1} \rangle \nonumber \\
    &~& +~\langle v^{12}_{12} \rangle + \langle v^{11}_{21} \rangle + \langle v^{12}_{11} \rangle + \langle v^{11}_{22} \rangle \nonumber \\
    &~& +~\langle v^{22}_{12} \rangle + \langle v^{21}_{22} \rangle + \langle v^{22}_{11} \rangle
\end{eqnarray}
The set of biorthogonal complements to $\{|\psi_{i_m}\rangle\}$ is denoted as
 $\{|\psi^{i_m}\rangle\}$. If
 the states \textit{within} each fragment are
 orthonormal, these
 are the same states, only built from the complement orbitals.
State indices have been suppressed on the right-hand side for brevity, and the symbol $\delta_m$ abbreviates $\delta_{i_m, j_m}$.
The angle-bracketed forms abbreviate summations to be defined presently, indicating the molecular integral involved and the 
fragments to which its respective indices are restricted.
For example,
\begin{eqnarray}
    \label{eq:h12}
    \langle t^{1}_{2} \rangle &=&
    (-1)^{n_{j_1}}  \sum_{p_1 q_2}    \rho_{p_1} \, \rho^{q_2} \, t^{p_1}_{q_2}
    \\
    \label{eq:h11}
    \langle {}^{2}u^{1}_{1} \rangle &=& 
    \sum_{\alpha_2 p_1 q_1} \rho_{p_1}^{q_1} \, {}^{\alpha_2}u^{p_1}_{q_1}
    \\
    \label{eq:v1212}
    \langle v^{12}_{12} \rangle &=&
    4 \sum_{p_1 q_2 r_1 s_2}   \rho_{p_1}^{r_1} \, \rho_{q_2}^{s_2} \, v^{p_1 q_2}_{r_1 s_2}
    \\
    \label{eq:v1121}
    \langle v^{11}_{21} \rangle &=&  
    2  (-1)^{n_{j_1}}  \sum_{p_1 q_1 r_2 s_1}    \rho_{p_1 q_1}^{s_1} \, \rho^{r_2} \, v^{p_1 q_1}_{r_2 s_1}
    \\
    \label{eq:v1122}
    \langle v^{11}_{22} \rangle &=&  
    \sum_{p_1 q_1 r_2 s_2}    \rho_{p_1 q_1} \, \rho^{s_2 r_2} \, v^{p_1 q_1}_{r_2 s_2}
\end{eqnarray}
A subscript on an index denotes restriction to the specified fragment, and $n_{i_m}$ is the number of electrons in state $|\psi_{i_m}\rangle$.
Using rules that will be addressed in Sec.~\ref{sec:workingequations}, the summations of molecular integrals and elements of density tensors $\boldsymbol{\rho}$ (which carry the implied state indices) on the right-hand side can be completely inferred from the abbreviations on the left.
All such expansions required for the numerical work in this article are recorded explicitly in Appendix \ref{sec:diagrams}.

The fact that the biorthogonal complements in a supersystem bra state can themselves be decomposed as a direct product of states associated with single fragments\cite{Dutoi.2019} is a crucial and nontrivial aspect that has allowed us to write these equations in terms of monomer transition-density tensors, which are themselves abbreviated according to the pattern
\begin{eqnarray}
    \rho_{p_m q_m}^{r_m}
    &=& \langle\psi^{i_m}| \hat{c}_{p_m} \hat{c}_{q_m} \hat{a}^{r_m} | \psi_{j_m}\rangle \\
    &=& \langle\psi^{i_m}| \hat{a}^\dagger_{p_m} \hat{a}^\dagger_{q_m} \hat{a}_{r_m} | \psi_{j_m}\rangle
\end{eqnarray}
with creation operators always preceding annihilation operators.
An extremely important aspect is that these density tensors are defined for the subsystems in isolation, meaning that we can suppress the machinery of the biorthogonal orbital representation in the second line, assuming the orbitals are orthonormal (and replace $\langle\psi^{i_m}|$ by $\langle\psi_{i_m}|$ if the fragment states are orthonormal).
It should also be pointed out that these tensors are zero unless the bra and ket have electron numbers that are commensurate with the field operators between them.
Computationally, eq.~\ref{eq:XRdimerelemtent} is better thought of as dividing into five overall-electron-conserving cases.
If no electrons are transferred between fragments 1 and 2, then only $\langle {}^{2}u^{1}_{1} \rangle$,  $\langle {}^{1}u^{2}_{2} \rangle$, and $\langle v^{12}_{12} \rangle$ contribute, whereas only $\langle v^{11}_{22} \rangle$ and $\langle v^{22}_{11} \rangle$ contribute to the cases of double charge transfers in opposite directions, and the remaining terms contribute only to the two single-charge-transfer cases.
Although this necessitates having both neutral and ionic basis states for each fragment, matrix elements for triple charge transfer and higher are zero, due to the two-electron nature of the \textit{ab initio} Hamiltonian.

With the separation just described, all information about how the fragments are arranged into a supersystem, and the consequences of overlapping fragment orbitals, are carried by the biorthogonalized molecular integrals,
leaving the density tensors effectively as per-fragment constants.
This is meant to modularize quantum chemistry using meaningful fragment quantities.
Reduced information about a set of high-quality states for each fragment (\textit{i.e.}, transition-density tensors) can be used as a permanent resource for supersystem calculations that rearrange them, computing the relevant molecular integrals as needed, per the geometric relationship between them. This algorithmically decouples the effort needed to solve for potentially intricate correlations within a fragment from the simpler interactions between them.
This
 is one of the core strengths of this methodology, which makes it stand out against other quantum chemical methods. In previous work, the existence of formulas like those in eqs.~\ref{eq:h12}--\ref{eq:v1122} was proven, but they had not yet been derived and implemented, a point we will return to shortly.

Eqs.~\ref{eq:abinitio} and \ref{eq:excitonic} are exactly equivalent Hamiltonians so long as all summations run over their complete ranges.  However, the purpose of rewriting the Hamiltonian in terms of fragments is to truncate it so that only a set of low-energy states for each fragment participate in the calculation, resulting in a large decrease in the size of the state space necessary to describe the couplings between molecules with accurate internal electronic structures.

Due to the biorthogonal representation, any truncated version of this operator is non-Hermitian, however.
One might expect that this formal deficiency would be practically unimportant for dealing with weaker intermolecular interactions, but, as will be discussed with the
 results
 here, once eqs.~\ref{eq:h12}--\ref{eq:v1122}, \textit{etc}., were fully derived and implemented, the effects of the truncation were found to be severe.
This is due to introducing unbalanced (\textit{i.e.}, non-Hermitian) terms of high energy, when the electrons of one molecule (\textit{e.g.}, in the core) try to use the orbitals of a neighboring molecule to increase their correlation. These discouraging results are now being presented alongside their proposed remedy.

\subsection{\label{sec:derivation} Development of a Hermitian Hamiltonian Expansion}

The encouraging numerical results
 previously published were obtained using an \textit{ad hoc} approximation, 
 in order
 to assess the viability of the XR
 concept. This was easier to implement than
 eqs.~\ref{eq:h12}--\ref{eq:v1122},
 \textit{etc}., but it would not be affordable for larger monomers.
The fact that this ``approximation'' gave better results than the eventually implemented parent theory
 can be traced to the fact that it is explicitly Hermitian.
This
 motivates putting it on its own theoretical footing here as part of a systematically improvable and computationally feasible scheme.

At the crux of the matter is the statement found already in the first article that ``the projected inverse of a matrix is not the same as the inverse of a projected matrix,'' which references the relationship between biorthogonal complements and the inverse of the overlap matrix of an original nonorthogonal set.\cite{Dutoi.2019} 
This statement highlights two well-defined choices of biorthogonal complements to the states $\{|\Psi_{\tilde{I}}\rangle\} \subset \{|\Psi_I\rangle\}$.
The tilde designates the \textit{model states} that span the \textit{model space}, per the aforementioned truncation that confines the allowed state indices $\tilde{i}_m$ of each fragment $m$ to a restricted set.

One choice of biorthogonal complements is $\{|\Psi^{\tilde{I}}\rangle\}$, which results from simply applying the same index restriction to the original full set of complements.
This is the ``projection of an inverse,'' making use of only a portion (projection) of the analytically known\cite{Dutoi.2019} complements to the complete set of states (inversion).
However $\{|\Psi_{\tilde{I}}\rangle\}$ and $\{|\Psi^{\tilde{I}}\rangle\}$ do not span the same space, which is the origin of the non-Hermiticity of the truncated Hamiltonian discussed above, leading to the poor results of the original theory that will be discussed below.

The other choice, developed in this work, would be the ``inverse of a projection,'' where we resolve the biorthogonal complements (inversion) only after limiting consideration to the model space (projection).
This alternate set of biorthogonal complements is a feature of the \textit{ad hoc} approximation that was used to good effect in our previous numerical work, although it was not presented in this language.
Under the designation XR$^\prime$, it was
 implemented there in a way that
 requires
 direct access to FCI wavefunctions for all dimers, a deficiency we now seek to remedy.

To obtain this alternate set of complements, let $\mathbf{S}$ be the matrix of overlaps for the model states only
\begin{eqnarray}
    \label{eq:Smatrix}
    \langle\Psi_{\tilde{I}}|\Psi_{\tilde{J}}\rangle = S_{\tilde{I}\tilde{J}} \in \mathbf{S}
\end{eqnarray}
and let us abbreviate its inverse as $\bar{\mathbf{S}}$
\begin{eqnarray}
    \label{eq:Sinverse}
    \bar{S}^{\tilde{I}\tilde{J}} \in \bar{\mathbf{S}} = \mathbf{S}^{-1}
\end{eqnarray}
which exists under the assumption that the set of model states is not linearly dependent.
We have here extended the meaning of the ``$\in$'' symbol to mean ``is an element of the matrix.''
We may now define the alternate set of biorthogonal complements via
\begin{eqnarray}
    \label{eq:newcomplements}
    \langle\bar{\Psi}^{\tilde{I}}| = \sum_{\tilde{J}} \bar{S}^{\tilde{I}\tilde{J}}\langle\Psi_{\tilde{J}}|
\end{eqnarray}
which satisfies $\langle\bar{\Psi}^{\tilde{I}}|\Psi_{\tilde{J}}\rangle = \delta_{\tilde{I}\tilde{J}}$ by definition and also manifestly spans the model space.
In addition to applying the above construction of complements to the overall supersystem, we can also apply it straightforwardly to its constituent subsystems (\textit{i.e.}, all possible monomers, dimers, trimers, \textit{etc.}), and this will be useful immediately below.

There are now two tasks ahead, in order to build this \textit{ad hoc}-but-promising method into one that is both systematically improvable and practically usable.  Taking the concerns in that order, we first need a rigorous analogue of eq.~\ref{eq:excitonic} using the new complements, and, second, we need to be able to compute the necessary matrix elements without recourse to anything more than molecular integrals and monomer transition-density tensors.

\subsubsection{\label{sec:hamiltonianexpansion} Hamiltonian Expansion}

Since the complements $\{|\bar{\Psi}^{\tilde{I}}\rangle\}$ lack the direct-product structure of $\{|\Psi^{\tilde{I}}\rangle\}$, we lose the simplicity of being able to immediately write formulas for  matrix elements in terms of ``substitutions'' (the number of fragments that have changed state between the bra and ket).\cite{Dutoi.2019}
Without this, we are forced to start from a full incremental expansion for the Hamiltonian operator, in parallel to the usual incremental expansion for the scalar energy.

Letting $\hat{\tilde{\mathcal{H}}}$ be the Hamiltonian in the model space, this reads
\begin{eqnarray}
    \label{eq:newexcitonic1}
    \hat{\tilde{\mathcal{H}}} &=&
    \tilde{H}_0 \nonumber \\
    &~& + \sum_{m} \sum_{\substack{\tilde{i}_m\\ \tilde{j}_m}} \tilde{H}^{\tilde{i}_m}_{\tilde{j}_m} \, \hat{\tau}^{\tilde{j}_m}_{\tilde{i}_m} \nonumber \\
    &~& + \sum_{m_1<m_2} \sum_{\substack{\tilde{i}_{m_1}, \tilde{i}_{m_2}\\ \tilde{j}_{m_1}, \tilde{j}_{m_2}}} \tilde{H}^{\tilde{i}_{m_1} \tilde{i}_{m_2}}_{\tilde{j}_{m_1} \tilde{j}_{m_2}} \, \hat{\tau}^{\tilde{j}_{m_1}}_{\tilde{i}_{m_1}} \hat{\tau}^{\tilde{j}_{m_2}}_{\tilde{i}_{m_2}} \nonumber \\
    &~& + \cdots \nonumber \\
    &~& + \sum_{\substack{\tilde{i}_{m_1}, \cdots \tilde{i}_{m_N}\\ \tilde{j}_{m_1}, \cdots \tilde{j}_{m_N}}} \tilde{H}^{\tilde{i}_{m_1} \cdots\tilde{i}_{m_N}}_{\tilde{j}_{m_1} \cdots \tilde{j}_{m_N}} \, \hat{\tau}^{\tilde{j}_{m_1}}_{\tilde{i}_{m_1}} \cdots \hat{\tau}^{\tilde{j}_{m_N}}_{\tilde{i}_{m_N}}
\end{eqnarray}
where the tildes on the scalar coefficients $\{\tilde{H}^{\tilde{i}_m}_{\tilde{j}_m}, \tilde{H}^{\tilde{i}_{m_1} \tilde{i}_{m_2}}_{\tilde{j}_{m_1} \tilde{j}_{m_2}}, \cdots \tilde{H}^{\tilde{i}_{m_1} \cdots\tilde{i}_{m_N}}_{\tilde{j}_{m_1} \cdots \tilde{j}_{m_N}}\}$ distinguish them from those in the original expansion of eq.~\ref{eq:excitonic}.
There is some freedom in choosing these coefficients.
For example, those pertaining to fluctuations of a specific fragment group can depend on reference states for spectator fragments outside that group.
We will here insist that $\hat{\tilde{\mathcal{H}}}$ has the same action on a subsystem (a monomer, dimer, \textit{etc.}) as the model-space Hamiltonian for that subsystem alone, \textit{i.e.}, the supersystem Hamiltonian directly contains all subsystem Hamiltonians.
This means that, beyond simply choosing fragment reference states to be the absolute vacuum for the electrons, their nuclei are not present either.
One immediate result of this choice is that the formally included reference energy $\tilde{H}_0$ is zero.

The full set of scalar coefficients for $\hat{\tilde{\mathcal{H}}}$ is then built from matrix elements of subsystem Hamiltonians according to the following pattern
\begin{eqnarray}
    \label{eq:newexcitonic2a}
    \tilde{H}_0
    &=& 0 
    \\
    \label{eq:newexcitonic2b}
    \tilde{H}^{\tilde{i}_{m}}_{\tilde{j}_{m}}
    &=& \langle\bar{\Psi}^{(\tilde{i}_{m})}|\hat{\mathcal{H}}_{\{m\}}| \Psi_{(\tilde{j}_{m})}\rangle - \tilde{H}_0 \, \delta_{\tilde{i}_{m} \tilde{j}_{m}} \\
    \label{eq:newexcitonic2c}
    \tilde{H}^{\tilde{i}_{m_1} \tilde{i}_{m_2}}_{\tilde{j}_{m_1} \tilde{j}_{m_2}}
    &=& \langle\bar{\Psi}^{(\tilde{i}_{m_1}, \tilde{i}_{m_2})}|
    \hat{\mathcal{H}}_{\{m_1,m_2\}}
    | \Psi_{(\tilde{j}_{m_1}, \tilde{j}_{m_2})}\rangle \nonumber \\
    &~& - \tilde{H}^{\tilde{i}_{m_1}}_{\tilde{j}_{m_1}} \, \delta_{\tilde{i}_{m_2} \tilde{j}_{m_2}} - \tilde{H}^{\tilde{i}_{m_2}}_{\tilde{j}_{m_2}} \, \delta_{\tilde{i}_{m_1} \tilde{j}_{m_1}} \nonumber \\
    &~& - \tilde{H}_0 \, \delta_{\tilde{i}_{m_1} \tilde{j}_{m_1}} \delta_{\tilde{i}_{m_2} \tilde{j}_{m_2}}
\end{eqnarray}
The bra states here are biorthogonal complements to the model states of the relevant subsystem only, containing no information about other fragments (\textit{e.g.}, via orbital biorthogonality).
Similarly, $\hat{\mathcal{H}}_\mathbf{m}$ denotes the complete Hamiltonian for the otherwise isolated system containing only the fragments present in $\mathbf{m}$.
We will henceforth suppress
 this
 subscript.
Consistent also with its use thus far, we will instead
 contextually interpret $\hat{\mathcal{H}}$ as being the
 full
 Hamiltonian for the
 isolated set of fragments
 whose state it is operating
 on.
A future broader formalism will build this behavior explicitly into the definition of $\hat{\mathcal{H}}$.

Finally, there is no guarantee of termination of eq.~\ref{eq:newexcitonic1} before the $N$-body term, though it is to be expected that interactions of many bodies are well-represented by the interactions among constituent groups, such that
high-order terms will not be routinely necessary.
In the limit of including all terms and considering all of the subtractions in eqs.~\ref{eq:newexcitonic2b}, \ref{eq:newexcitonic2c}, \textit{etc.}, we clearly have
$\hat{\tilde{\mathcal{H}}} = \hat{P}_\perp \hat{\mathcal{H}} \hat{P}_\perp$
where $\hat{P}_\perp =\sum_{\tilde{I}} |\Psi_{\tilde{I}}\rangle\langle\bar{\Psi}^{\tilde{I}}|$ is the orthogonal projector onto the supersystem model space.
It is noteworthy then that the application of the original truncated set of full complements $\{|\Psi^{\tilde{I}}\rangle\}$ to the incremental expansion would result in an oblique projection of the Hamiltonian $\hat{P}_\angle \hat{\mathcal{H}} \hat{P}_\angle$,
with $\hat{P}_\angle =\sum_{\tilde{I}} |\Psi_{\tilde{I}}\rangle\langle\Psi^{\tilde{I}}|$,
 because $\{|\Psi_{\tilde{I}}\rangle\}$ and $\{|\Psi^{\tilde{I}}\rangle\}$ do not span the same space.
The result of this is, however, completely redundant with simply truncating eq.~\ref{eq:excitonic}, on account of the subtractions in eqs.~\ref{eq:newexcitonic2b}, \ref{eq:newexcitonic2c}, \textit{etc.}, and the previously derived substitution rules.\cite{Dutoi.2019}
There is a rich set of mathematical connections to explore, starting with the mechanism for the mandatory disappearance of high-order terms as the number of fragment states increases ($\hat{P}_\angle\rightarrow\hat{P}_\perp\rightarrow 1$), which can be itself understood in terms of a fragment cluster expansion.
This will
 also
 be the subject of future work.

\subsubsection{\label{sec:matrixelements} Matrix Elements}

Since the bra in the matrix elements in eqs.~\ref{eq:newexcitonic2a}--\ref{eq:newexcitonic2c} cannot be decomposed in tensor-product form by fragment, we must apply the formulas for numerically resolved biorthogonal complements in terms of the overlap matrix, resulting in
\begin{eqnarray}
    \label{eq:matrixelements}
    \langle\bar{\Psi}^{\tilde{I}}|\hat{\mathcal{H}}| \Psi_{\tilde{J}}\rangle
    ~=~ [\bar{\mathbf{S}}\tilde{\mathbf{H}}]_{\tilde{I}\tilde{J}}
\end{eqnarray}
where the right-hand side denotes an element of a matrix product, and
\begin{eqnarray}
    \label{eq:Hmatrix}
    \tilde{H}_{\tilde{I}\tilde{J}} 
    = \langle\Psi_{\tilde{I}}|\hat{\mathcal{H}}| \Psi_{\tilde{J}}\rangle \in \tilde{\mathbf{H}}
\end{eqnarray}
is an element of the Hamiltonian in the symmetric, non-orthogonal representation for the relevant group of fragments. Although $\bar{\mathbf{S}}\tilde{\mathbf{H}}$ is not Hermitian, it is a complete representation of the action of $\hat{\mathcal{H}}$ within the model space for those fragments, and contains no information from outside the model space. It is related by a similarity transformation to the manifestly Hermitian matrix $\bar{\mathbf{S}}^{1/2}\tilde{\mathbf{H}}\bar{\mathbf{S}}^{1/2}$, having real eigenvalues and a variational lowest energy.

It can be observed that eqs.~\ref{eq:newexcitonic2c} and \ref{eq:matrixelements} describe directly the treatment of dimer couplings in the prior \textit{ad hoc} numerical work.\cite{Liu.2019}
In that work, $\mathbf{S}$ and $\tilde{\mathbf{H}}$ for each dimer were built using direct access to the underlying small-system/small-basis FCI calculation for that dimer.
This is clearly not a scalable strategy for larger monomers, and it is at odds with the advertised advantages of eqs.~\ref{eq:h12}--\ref{eq:v1122}, and analogues.
The abstract equations for these matrix elements have now been expounded upon in the language of an alternative set of biorthogonal complements, which we will presently make use of to try to repair this deficiency, in order to make a scalable method.

We must now compute the matrices $\mathbf{S}$ and $\tilde{\mathbf{H}}$ and invert $\mathbf{S}$.
Since the principle underlying XR methods is that only relatively small numbers of states should be needed for each fragment, and only relatively small groups of fragments need to be considered in building the supersystem Hamiltonian, we assert that it will be sufficient to numerically invert $\mathbf{S}$ with general-purpose libraries and well-studied techniques.
The more formidable difficulty is obtaining these matrices in the first place.
For this, we will begin with $\mathbf{S}$ and then generalize the discussion to $\tilde{\mathbf{H}}$.

In computing $\mathbf{S}$, we are working now in the symmetric representation, where both the bra and ket can be decomposed by fragment in tensor-product form
\begin{eqnarray}
    \label{eq:covarianttenspdt}
    S_{(\tilde{i}_{m_1}, \tilde{i}_{m_2}, \cdots),(\tilde{j}_{m_1}, \tilde{j}_{m_2}, \cdots)} ~~~~~~~~~~~~~~~~~~~~~~~~~~~~\nonumber \\
    ~=~
    \langle \psi_{\tilde{i}_{m_1}}\psi_{\tilde{i}_{m_2}}\cdots | \psi_{\tilde{j}_{m_1}}\psi_{\tilde{j}_{m_2}}\cdots\rangle
\end{eqnarray}
Theoretically, this could be evaluated using the standard formula for determinant overlaps between every pair of configurations comprising general bra and ket states.
However, such an approach would demand keeping complete wavefunctions for the fragments and have prohibitive computational scaling.

Our approach will be to derive an expansion for $\mathbf{S}$, where the leading-order term for each element is just a product of Kronecker deltas for the fragment states.
This effectively presumes orthonormality of all orbitals at zeroth order, along with the permanent assumption that states within each fragment are orthonormal (the latter will be addressed shortly).
Understanding that the exact formula for a many-electron overlap is an antisymmetrized product of orbital overlaps, increasing levels of approximation should include higher-order products of interfragment orbital overlaps, the vast majority of which will be vanishingly small.
The basic ingredients then are orbital overlaps and also field operators to probe occupations and coherences.

To make connection to the orbitals we first turn to the basis of supersystem determinants $\{|\Phi_P\rangle\}$.
Again, following prior notation, we have 
\begin{eqnarray}
    \label{eq:compositestates}
    |\Psi_{I}\rangle
    &=& \sum_{P} |\Phi_{P}\rangle Z^{P}_{I}
\end{eqnarray}
where $P$ is a tuple that lists ``occupied'' orbital indices in ascending order (according to some pre-defined fragment-blocked list), and which can be broken into sub-tuples $P_1, \cdots P_N$, containing the orbital indices for each fragment.
A supersystem determinant can then also be expressed in terms of determinant states $\{|\phi_{P_m}\rangle\}$ for each fragment $m$ as
\begin{eqnarray}
    |\Phi_P\rangle
    ~=~ |\phi_{P_1} \cdots \phi_{P_N}\rangle 
\end{eqnarray}
This also provides
\begin{eqnarray}
    Z^{P}_{I}
    &=& \prod_{m} z^{P_m}_{i_m}
\end{eqnarray}
in terms of the coefficients that describe the electronically correlated fragment states
\begin{eqnarray}
    \label{eq:fragmentstates}
    |\psi_{i_m}\rangle
    &=& \sum_{P_m} |\phi_{P_m}\rangle z^{P_m}_{i_m}
\end{eqnarray}
$Z^{P}_{I}$$\in$$\mathbf{Z}$ and $z^{P_m}_{i_m}$$\in$$\mathbf{z}$ abstractly denote invertible square matrices of wavefunction coefficients,
though only \textit{indirect} knowledge of a small fraction of $\mathbf{z}$, which is block-diagonal by fragment and electron number, will ever be needed in practice.
Finally, $\{|\Phi^P\rangle\}$ and $\{|\phi^{P_m}\rangle\}$ denote the sets of biorthogonal complements to the original determinant bases ($\langle\Phi^P|\Phi_Q\rangle = \delta_{PQ}$ and $\langle\phi^{P_m}|\phi_{Q_m}\rangle = \delta_{P_m Q_m}$), built by making determinants of orbitals from the complement set $\{|\chi^p\rangle\}$.

Since $|\Phi^P\rangle$ has the same information content as $|\Phi_P\rangle$ (one-to-one mapping), we can reason that a path forward is to seek an operator $\hat{S}$ that satisfies the following in the biorthogonal representation
\begin{eqnarray}
    \label{eq:detoverlap}
    \langle \Phi_P|\Phi_Q\rangle = \langle \Phi^P | \hat{S} | \Phi_Q\rangle
\end{eqnarray}
A leading order term of unity for $\hat{S}$ would agree with our stated goal,
and working in the biorthogonal representation will allow us to resolve the rest of the expansion for $\hat{S}$ in terms of field operators with canonical anticommutation relationships.
If $\hat{S}$ can be written as just proposed, then we know from previous arguments,\cite{Dutoi.2019} that the evaluation can be carried through in terms of single-fragment transition-density tensors.
Furthermore, a field-operator-based formulation makes it trivially generalizable to arbitrary fragment-decomposed states in the bra and ket, simply requiring transition-density tensors for general fragment states, consistent with our overall theoretical framework.

Now, we would like to also write
\begin{eqnarray}
    \label{eq:targetoverlap}
    \langle \Psi_I|\Psi_J\rangle = \langle \Psi^I | \hat{S} | \Psi_J\rangle
\end{eqnarray}
for an element of $\mathbf{S}$ (for model-state indices).
But this is only true if $\mathbf{z}$ (and therefore $\mathbf{Z}$) is unitary, since the definition of $\langle\Psi_I|$ involves conjugating $\mathbf{Z}$ and the definition of $\langle\Psi^I|$ involves inverting it.
For convenience, we will make this assumption that $\mathbf{z}^\dagger = \mathbf{z}^{-1}$.
This is safe in the case that the overlap matrix is anyway computable for the many-electron states of a single fragment because an orthonormal single-fragment basis $\{|\psi_{i_m}\rangle\}$ can be constructed from a non-orthonormal precursor for no more work than what is involved in constructing the monomer Hamiltonian terms according to eq.~\ref{eq:newexcitonic2b}.
If a non-Hermitian method is employed for the fragments (like equation-of-motion coupled-cluster, EOM-CC), making the single-fragment overlap matrix practically uncomputable, there is still an exact rewriting of the Hamiltonian that takes advantage of the present developments, but it will not be overall Hermitian.
This generalization is discussed in Appendix \ref{sec:nonhermitianfrags}.

An abstract writing for $\hat{S}$ is readily available as
\begin{eqnarray}
    \label{eq:abstractS}
    \hat{S} = \sum_{P} |\Phi_P\rangle\langle\Phi_P|
\end{eqnarray}
revealing its nature as a Hermitian operator that transforms between the two orbital representations for  determinant states according to
\begin{eqnarray}
    \label{eq:Sforwardaction}
    |\Phi_P\rangle = \hat{S} |\Phi^P\rangle
\end{eqnarray}
Whereas eq.~\ref{eq:abstractS} can be convenient for abstract work, for working equations, it is more useful to start by noting that satisfying  eq.~\ref{eq:Sforwardaction} for arbitrary $|\Phi^P\rangle$ also fully defines $\hat{S}$.
It is essentially an orbital-replacement operator, and it can be verified that eq.~\ref{eq:Sforwardaction} is also satisfied by
\begin{eqnarray}
    \label{eq:verifySaction}
    \hat{S}
    ~=~ 1&+&
    \sum_p (\hat{c}_p-\hat{c}^p)\hat{a}_p  \nonumber \\
    &+&\sum_{p<q} (\hat{c}_p-\hat{c}^p) (\hat{c}_q-\hat{c}^q)\hat{a}_q \hat{a}_p  ~+~ \cdots \\
    \label{eq:Spractical}
    ~=~ 1 &+& \sum_{p q} \sigma_{p q} \, \hat{c}_p \hat{a}^q \nonumber \\
    &+& \frac{1}{2} \sum_{p q r s} \sigma_{p q} \sigma_{r s} \, \hat{c}_p \hat{c}_r \hat{a}^s \hat{a}^q ~+~ \cdots \\
    \label{eq:Sorders}
    ~=~ \hat{S}^{[0]} &+& \hat{S}^{[1]} ~+~ \hat{S}^{[2]} ~+~ \cdots
\end{eqnarray}
which references the deviation of the orbital overlaps from the unit matrix
\begin{eqnarray}
    \sigma_{pq}
    &\in& \boldsymbol{\sigma} ~=~ \mathbf{s} - \mathbf{1} \\
    s_{pq} &=& \langle \chi_p | \chi_q \rangle ~\in~ \mathbf{s}
\end{eqnarray}
Eq.~\ref{eq:verifySaction} is useful for verifying that this is equivalent to $\hat{S}$ in eq.~\ref{eq:Sforwardaction}, by applying complete induction to the resolutions of its action onto  determinants of orbitals from $\{|\chi^p\rangle\}$ with successively higher electron number, starting with the absolute vacuum $|\rangle$.
Eq.~\ref{eq:Spractical} will be practical for deriving working equations in canonical form.
The equivalence between eqs.~\ref{eq:verifySaction} and 
\ref{eq:Spractical} requires first recalling that $\hat{S}$ is Hermitian and working with the conjugate of the right-hand side of eq.~\ref{eq:verifySaction}.
Then we can use
\begin{eqnarray}
    \hat{a}_p &=& \sum_q s_{pq} \hat{a}^q
\end{eqnarray}
which, along with similar relationships between $\hat{c}_p$ and $\hat{c}^p$, can be inferred from the relationship of the biorthogonal single-electron bases to the orbital overlaps. Kronecker deltas that result from reordering field operators are then absorbed into the definition of $\boldsymbol{\sigma}$.
Eq.~\ref{eq:Sorders} simply provides a nomenclature for the terms of different order in $\boldsymbol{\sigma}$ in eq.~\ref{eq:Spractical}.

It is satisfying that the deviation of $\mathbf{S}$ from matrix unity in the many-electron space (eqs.~\ref{eq:detoverlap} and \ref{eq:targetoverlap}) has a direct relationship to the deviation of $\mathbf{s}$ from matrix unity in the orbital space.
Finally, it is interesting that representation of $\hat{S}$ as a term-by-term reordering of $\text{e}^{\hat{\sigma}}$ ($\hat{\sigma} = \sum_{pq} \sigma_{p q} \, \hat{c}_p \hat{a}^q$) highlights the nature of overlaps of many-electron states as antisymmetrized products of orbital overlaps.
It also suggests a rapidly converging series since the values in $\boldsymbol{\sigma}$ are all less than 1.

The forgoing has now accomplished the separation that we seek.
Inserting eq.~\ref{eq:Spractical} into eq.~\ref{eq:targetoverlap} gives an expansion for the target overlaps, where each term is a product of orbital overlap information (via elements of $\boldsymbol{\sigma}$) and transition-density tensors (via matrix elements of field-operator strings).
The latter half of this claim will be scrutinized more carefully as we derive working equations in the next section.

Lastly, before deriving working equations, we discuss the generalization to
\begin{eqnarray}
    \label{eq:targetH}
    \langle \Psi_I|\hat{\mathcal{H}}|\Psi_J\rangle = \langle \Psi^I | \hat{S} \hat{\mathcal{H}}| \Psi_J\rangle
\end{eqnarray}
for an element of $\tilde{\mathbf{H}}$ (for model-state indices).
These have the same structure as those for $\mathbf{S}$, and so they will also have the same advertised decomposition.
Inserting eqs.~\ref{eq:abinitio} and \ref{eq:Spractical} into the right-hand side simply generates terms whose scalar coefficients are products of molecular integrals and elements of $\boldsymbol{\sigma}$, and the matrix elements of field-operator strings will nevertheless be resolved in terms of transition-density tensors for monomers, as we shall soon see.

To facilitate the writing of eq.~\ref{eq:targetH} in terms of transition-density tensors, it will be useful to first reorder the field operators of $\hat{S}\hat{\mathcal{H}}$ so that creation operators are always to the left of annihilation operators.
Term-by-term application of the canonical anticommutation relationships and recognition that the result can be written again compactly in terms of $\hat{S}$ yields
\begin{eqnarray}
    \label{eq:BOleadingterm}
    \hat{S}\hat{\mathcal{H}} = \widehat{S\mathcal{H}}^{(0)} + \widehat{S\mathcal{H}}^{(1)} + \widehat{S\mathcal{H}}^{(2)} 
\end{eqnarray}
where
\begin{eqnarray}
    \label{eq:commutatorS0}
    \widehat{S\mathcal{H}}^{(0)}
    &=& \sum_{pqrs} h^p_q \, \hat{c}_p \hat{S} \hat{a}^q + \sum_{pqrs} v^{pq}_{rs} \, \hat{c}_p \hat{c}_q \hat{S} \hat{a}^s \hat{a}^r \\
    \label{eq:commutatorS1}
    \widehat{S\mathcal{H}}^{(1)}
    &=& \sum_{pqrs} (h_{pq} - h^p_q) \, \hat{c}_p \hat{S} \hat{a}^q \\
    &~& +
    \sum_{pqrs} (2v^p_{qrs} - 2v^{pq}_{rs}) \, \hat{c}_p \hat{c}_q \hat{S} \hat{a}^s \hat{a}^r \nonumber \\
    \label{eq:commutatorS2}
    \widehat{S\mathcal{H}}^{(2)}
    &=& \sum_{pqrs} (v_{pqrs} - 2v^p_{qrs} + v^{pq}_{rs}) \, \hat{c}_p \hat{c}_q \hat{S} \hat{a}^s \hat{a}^r
\end{eqnarray}
with
\begin{eqnarray}
    \label{eq:symmh}
    h_{pq} &=& \langle\chi_p|\hat{h}|\chi_q\rangle \\
    \label{eq:halfv}
    v^p_{qrs} &=& \frac{1}{2}\big[
    \frac{1}{4}\langle\chi^p\chi_q|\hat{v}|\chi_r\chi_s\rangle + 
    \frac{1}{4}\langle\chi_p\chi^q|\hat{v}|\chi_r\chi_s\rangle
    \big] \\
    \label{eq:symmv}
    v_{pqrs} &=& \frac{1}{4}\langle\chi_p\chi_q|\hat{v}|\chi_r\chi_s\rangle
\end{eqnarray}
where we have used the common notation for the combined kinetic and nuclear attraction operator $\hat{h} = \hat{t} + {}^{\mathbf{m}}\hat{u}$, where the set of fragments in the left-hand superscript indicates summation of ${}^{\alpha}\hat{u}$ over the atoms belonging to those fragments.
The parenthesized superscripts in eqs.~\ref{eq:BOleadingterm}--\ref{eq:commutatorS2} are the numbers of Kronecker deltas from the anticommutations that have been resolved to produce each term, which is limited to two because of the two-electron Hamiltonian.
These Kronecker deltas have the effect of contracting elements of $\boldsymbol{\sigma}$ with molecular integrals, effectively changing their bases.

The reason for writing the expansion in three pieces in eq.~\ref{eq:BOleadingterm} is to highlight a nuance in keeping track of orders of $\boldsymbol{\sigma}$.
$\hat{S}^{[0]}\hat{\mathcal{H}}$ is equivalent to expanding $\hat{S}$ in the definition of $\widehat{S\mathcal{H}}^{(0)}$ to zeroth order, which we will now denote $\widehat{S\mathcal{H}}^{(0)[0]}$.
In general we have
\begin{eqnarray}
    \label{eq:SH0}
    \hat{S}^{[0]}\hat{\mathcal{H}} &=& \widehat{S\mathcal{H}}^{(0)[0]} \\
    \label{eq:SH1}
    \hat{S}^{[1]}\hat{\mathcal{H}} &=& \widehat{S\mathcal{H}}^{(0)[1]} + \widehat{S\mathcal{H}}^{(1)[0]} \\
    \label{eq:SH2}
    \hat{S}^{[2]}\hat{\mathcal{H}} &=& \widehat{S\mathcal{H}}^{(0)[2]} + \widehat{S\mathcal{H}}^{(1)[1]} + \widehat{S\mathcal{H}}^{(2)[0]} \\
    \label{eq:SH3}
    \hat{S}^{[3]}\hat{\mathcal{H}} &=& \widehat{S\mathcal{H}}^{(0)[3]} + \widehat{S\mathcal{H}}^{(1)[2]} + \widehat{S\mathcal{H}}^{(2)[1]}
\end{eqnarray}
and so forth.
The first term on the right-hand side in each of these is transparently of the same order in $\boldsymbol{\sigma}$ as expected from the notation on the left-hand side, but the further terms have powers of $\boldsymbol{\sigma}$ ``hidden'' in the transformations of the integrals.

Adding together the three series in eq.~\ref{eq:BOleadingterm} results in 
\begin{eqnarray}
    \label{eq:termwisehermitian}
    \hat{S}\hat{\mathcal{H}}
    &=& \sum_{pqrs} h_{pq} \, \hat{c}_p \hat{S} \hat{a}^q + \sum_{pqrs} v_{pqrs} \, \hat{c}_p \hat{c}_q \hat{S} \hat{a}^s \hat{a}^r
\end{eqnarray}
which has a pleasantly simple structure, but it should be pointed out that the terms that result by expanding $\hat{S}$ on the right-hand side of this are of mixed order in $\boldsymbol{\sigma}$, relative to eqs.~\ref{eq:SH0}--\ref{eq:SH3}.
The two series can be thought of as expansions around different points in operator space.
The zeroth order term of eq.~\ref{eq:SH0} already incorporates some overlap effects through the biorthogonalized molecular integrals, whereas the zeroth-order term of eq.~\ref{eq:termwisehermitian} treats the fragments as if the orbitals were non-overlapping even as they are already physically interacting.
The former will be seen to have better convergence properties, but the latter (using the symmetric integrals) shows that, as is true for $\mathbf{S}$, our expressions converge to a Hermitian matrix for $\tilde{\mathbf{H}}$.
It is also worthwhile to note that eq.~\ref{eq:termwisehermitian} can be derived more elegantly by starting with an alternate form of the \textit{ab initio} Hamiltonian (written in terms of field operators that do not obey canonical anticommutation relationships)
\begin{eqnarray}
    \hat{\mathcal{H}}
    &=& \sum_{pqrs} h_{pq} \hat{c}^p \hat{a}^q + \sum_{pqrs} v_{pqrs} \hat{c}^p \hat{c}^q \hat{a}^s \hat{a}^r 
\end{eqnarray}
and applying the identity
\begin{eqnarray}
    \hat{c}_p \hat{S} = \hat{S} \hat{c}^p
\end{eqnarray}
which can be inferred from eq.~\ref{eq:Sforwardaction}.

\subsection{\label{sec:workingequations}Working Equations}

Here we outline the procedure for deriving working equations for the elements of $\mathbf{S}$ and $\tilde{\mathbf{H}}$ that are necessary to build the Hamiltonian expansion of 
 eq.~\ref{eq:newexcitonic1} (via eqs.~\ref{eq:newexcitonic2a}--\ref{eq:matrixelements}, \ref{eq:targetoverlap} \& \ref{eq:targetH}).
More detail is given in the context of an explicitly derived example in Appendix \ref{sec:work_eq_ex}.
In general, we must resolve quantities of the form
\begin{eqnarray}
    \label{eq:fieldopsandwich}
    \langle \Psi^I | \hat{c}_p \hat{c}_q \cdots \hat{a}^r \hat{a}^s | \Psi_J \rangle ~~~~~~~~~~~~~~~~~~~~~~~~~~~~~~~~~~~~~~~ \nonumber \\
    =~ \sum_{P_1} \cdots \sum_{P_N} \sum_{Q_1} \cdots \sum_{Q_N} z_{i_1}^{{P_1}*}\cdots z_{i_N}^{{P_N}*} z_{j_1}^{Q_1}\cdots z_{j_N}^{Q_N} \nonumber \\
    \langle \hat{\phi}^{P_N} \cdots \hat{\phi}^{P_1} \hat{c}_p \hat{c}_q \cdots \hat{a}^r \hat{a}^s \hat{\phi}_{Q_1} \cdots \hat{\phi}_{Q_N} \rangle
\end{eqnarray}
where the angle brackets indicate a vacuum expectation value, and the operators $\hat{\phi}_{P_m}$ and $\hat{\phi}^{P_m}$ are nothing more than strings of field operators that satisfy
\begin{eqnarray}
    |\phi_{P_m}\rangle = \hat{\phi}_{P_m}|\rangle \\
    \langle \phi^{P_m}| = \langle|\hat{\phi}^{P_m}
\end{eqnarray}
Using the canonical anticommutation rules, the expectation value of the string of field operators for all fragments can be separated into a product of expectation values for the separate fragments,
and any accumulated phases are global for the summation, as long as the states $|\psi_{i_m}\rangle$ individually have well-defined electron number.
Briefly, since operators on different fragments simply anticommute, after reordering by fragment, each fragment string can be rewritten as the sum of a scalar and normal-ordered strings, leaving only a product of the scalars for each fragment.
These can then be recontracted with the elements of $\mathbf{z}$ in eq.~\ref{eq:fieldopsandwich}, meaning that knowledge of quantities of the form
\begin{eqnarray}
    \label{eq:transdensities}
    \langle \psi^{i_m}|\hat{c}_{p_m} \cdots \hat{a}^{q_m} | \psi_{j_m} \rangle
    &=&
    \langle \psi_{i_m}|\hat{a}^\dagger_{p_m} \cdots \hat{a}_{q_m} | \psi_{j_m} \rangle \\
    &=& \rho_{p_m \cdots}^{\cdots q_m}
\end{eqnarray}
for each fragment is sufficient.
The reference to biorthogonalized orbitals can be suppressed for a fragment in isolation, since the canonical anticommutation relationships are all that is needed to evaluate these.
We can also replace $|\psi^{i_m}\rangle$ by $|\psi_{i_m}\rangle$ in this case, since we have asserted that $\mathbf{z}^\dagger = \mathbf{z}^{-1}$.
These quantities are the transition-density tensors to which we have previously referred.
There is no requirement that the electron count be the same in both the bra and ket for a given fragment, only that the total number of electrons is conserved for all fragments
 in a supersystem matrix element of $\hat{S}$ or $\hat{S}\hat{\mathcal{H}}$.

The remaining effort is now the application of the foregoing workflow to each of the terms that arise; for different operators, at different orders in $\boldsymbol{\sigma}$, for different numbers of fragments, and for the various subcases within each such summation that distributes orbital indices differently over the fragments.
Since eqs.~\ref{eq:Spractical}, \ref{eq:commutatorS0}--\ref{eq:commutatorS2} \& \ref{eq:termwisehermitian} differ only in the scalar coefficients of their respective operator expansions it suffices for us to know how to build general matrix elements of $\hat{S}^{[o]}$, $\sum_{pq} h^p_q \, \hat{c}_p \hat{S}^{[o]} \hat{a}^q$, and $\sum_{pqrs} v^{pq}_{rs} \, \hat{c}_p \hat{c}_q \hat{S}^{[o]} \hat{a}^s \hat{a}^r$ for arbitrary order $o$.
Focusing, for example, on the two electron part of a dimer matrix element, we have, for $o$$=$$0$ 
\begin{eqnarray}
    \label{eq:explicitmatrixelementA}
    \langle\psi^{i_1} \psi^{i_2} |
      \Big[ \frac{1}{0!}
        \sum_{pqrs} v^{pq}_{rs} \, \hat{c}_p \hat{c}_q  \hat{a}^s \hat{a}^r
      \Big]
    |\psi_{j_1} \psi_{j_2}\rangle
    = ~~~~~~~~~~~~~~~ \\
      \langle v^{11}_{11} \rangle\delta_2
    + \langle v^{22}_{22} \rangle\delta_1
    + \langle v^{12}_{12} \rangle ~~~~~~~~~~~~~~~~~~~~~~~~~~~~~~~
    \nonumber \\
    +~\langle v^{11}_{21} \rangle
    + \langle v^{12}_{11} \rangle
    + \langle v^{11}_{22} \rangle
    + \langle v^{22}_{12} \rangle
    + \langle v^{21}_{22} \rangle
    + \langle v^{22}_{11} \rangle \nonumber
\end{eqnarray}
where we have already encountered, in eqs.~\ref{eq:h12}--\ref{eq:v1122}, abbreviations of the type used for the terms on the right-hand side, with state indices again suppressed.
Each such summation has the same structure as the left-hand side, but with orbital indices restricted to specific fragments.
They also account for equivalent terms arising from permutational (anti)symmetries; with $4$ indices that can each run over the orbitals of two fragments, there are naively $2^4$$=$$16$ cases of such restricted summations, but only $9$ numerically unique terms appear in eq.~\ref{eq:explicitmatrixelementA}.
The prefactor of $2$ in eq.~\ref{eq:v1121} for $\langle v^{11}_{21} \rangle$ results from the fact that two of the \textit{a priori} $16$ cases evaluate to the summation that it multiplies.
Furthermore, the phase depends on the number of electrons in $|\psi_{j_1}\rangle$ because the charge-transfer character of this element causes us to permute strings with odd numbers of field operators in the course of collecting all operators together by fragment when building the vacuum expectation value of eq.~\ref{eq:fieldopsandwich}.
If all indices lie on a single fragment, then we obtain a Kronecker delta with respect to the state of the other fragment due to the assumption that the single-fragment states are orthonormal.
We include the explicit denominator of $o!$ in eq.~\ref{eq:explicitmatrixelementA} to make clear that these exponential-like expansion
 coefficients
 are
 part of the definitions of the $\hat{S}^{[o]}$
 terms
 and are folded into
 the definitions of the abbreviations for the sums just discussed.

For convenient identification, these angle-bracket-abbreviated terms will be called \textit{diagrams} because graphs representing fragments as nodes and integrals as directional edges (multipartite, in the case of $v^{pq}_{rs}$) have proved useful for tracking the larger number of them when more fragments are involved.
A full presentation of this would not be justified for only the dimer terms in this article, and it will be postponed for a future article on implementation.

Finally, let us introduce an example of a symbol for a diagram that contains $\boldsymbol{\sigma}$ by way of writing the two-electron part of a dimer matrix element for $o$$=$$1$ 
\begin{eqnarray}
    \label{eq:explicitmatrixelementB}
    \langle\psi^{i_1} \psi^{i_2} |
      \Big[ \frac{1}{1!}
        \sum_{pqrstu} \sigma_{tu} v^{pq}_{rs} \, \hat{c}_p \hat{c}_q  \hat{c}_t  \hat{a}^u \hat{a}^s \hat{a}^r
      \Big]
    |\psi_{j_1} \psi_{j_2}\rangle
    ~=~ ~~~~~~~~ \\
      \langle \sigma_{12} v^{12}_{11} \rangle
    + \langle \sigma_{12} v^{22}_{21} \rangle
    + \langle \sigma_{12} v^{12}_{12} \rangle
    + \langle \sigma_{12} v^{11}_{11} \rangle
    + \langle \sigma_{12} v^{22}_{22} \rangle \nonumber\\
    +~\langle \sigma_{12} v^{22}_{11} \rangle
    + \langle \sigma_{12} v^{11}_{21} \rangle
    + \langle \sigma_{12} v^{12}_{22} \rangle
    + \langle \sigma_{12} v^{11}_{22} \rangle \nonumber\\
    +~\langle \sigma_{21} v^{21}_{22} \rangle
    + \langle \sigma_{21} v^{11}_{12} \rangle
    + \langle \sigma_{21} v^{21}_{21} \rangle
    + \langle \sigma_{21} v^{22}_{22} \rangle
    + \langle \sigma_{21} v^{11}_{11} \rangle \nonumber\\
    +~\langle \sigma_{21} v^{11}_{22} \rangle
    + \langle \sigma_{21} v^{22}_{12} \rangle
    + \langle \sigma_{21} v^{21}_{11} \rangle
    + \langle \sigma_{21} v^{22}_{11} \rangle \nonumber
\end{eqnarray}
The explicit forms of the summations needed to evaluate these terms, along with all other diagrams needed to obtain the numerical results in this work are listed in Appendix \ref{sec:work_eqs}.
In doing this, we can safely assume that all elements of $\boldsymbol{\sigma}$ with both indices on the same fragment are zero by assuming that the orbitals within any fragment are orthonormal.
It is trivial to transform any transition-density tensor to a basis that conforms to this assumption.

\subsection{\label{sec:nomenclature}Method Nomenclature}

A new class of methods is now introduced based on the order-by-order expansion of $\hat{S}$.
This occurs in two places in approximating eq.~\ref{eq:matrixelements} for each fragment group in the excitonic Hamiltonian.
It appears once in building $\mathbf{S}$ itself (which is numerically inverted), and once in building the symmetric nonorthogonal resolution of the Hamiltonian $\tilde{\mathbf{H}}$.
To be internally consistent, $\hat{S}$ is approximated to the same order in each place
(this also appears to be mandatory to obtain reasonable results).
Once these matrix elements are computed, the excitonic Hamiltonian of eq.~\ref{eq:newexcitonic1} can be built via eqs.~\ref{eq:newexcitonic2a}--\ref{eq:newexcitonic2c}, \textit{etc.}, and subjected to an independently chosen algorithm for further simulations, such as finding the ground state energy.

The method nomenclature is now given as XR, followed by an integer, denoting the fragment order at which
 eq.~\ref{eq:newexcitonic1} is truncated, and a bracketed integer at the end, indicating the maximum order considered in the expansion of $\hat{S}$ according to eq.~\ref{eq:Sorders}
 (\textit{i.e.}, the order of the interfragment orbital overlaps $\boldsymbol{\sigma}$). 
For instance XR2[1] refers to an XR Hamiltonian including up to dimer interactions and first-order terms in $\hat{S}$.
However, the fragment order indicator may be suppressed when it is either clear or irrelevant to the discussion at hand.

Without the order for $\hat{S}$ indicated in a method identifier, the original XR theory is implied, using the (truncated) Hamiltonian of eq.~\ref{eq:excitonic}, as opposed to eq.~\ref{eq:newexcitonic1}.
In sentence form, the abbreviation XR may still refer to all of excitonic renormalization theory in general, and the phrase ``XR proper'' will be reserved to specify a method based on the original, globally biorthogonal machinery.
XR proper is distinct from XR[0] because the states used in the monomer terms of XR[0] have absolutely no knowledge of other fragments, whereas they are affected by the global biorthogonalization in XR proper (as discussed immediately after
 eqs.~\ref{eq:newexcitonic2a}--\ref{eq:newexcitonic2c}).
However, for fragment-complete Hamiltonians (\textit{e.g.}, including up to three-body couplings for trimer systems, \textit{etc.}) there is no numerical difference between XR and XR[0], since they amount to alternate partitionings of the full zeroth order Hamiltonian.
It will be interesting in the future to explore the connection, and perhaps method space, between XR and XR[0].
The limit of this series, XR[$\infty$], can be computed by brute force using monomer FCI wavefunctions, and has already been shown (under the designation XR$^\prime$) to be effectively exact for dimers, even with sharply truncated state spaces for the atoms.\cite{Dutoi.2019}

This now fulfills one of the aims of this work, which is to put the promising results from the previous \textit{ad hoc} approximation on solid formal footing.
This effectively exact method is now expressed as the limit of a series, which can be approached stepwise.
It is the convergence of this series that we now explore, providing evidence that it is quick. Hence, the newly developed XR family of methods shows great potential for becoming practically relevant.

\section{\label{sec:results} Results and Discussion}

\subsection{\label{sec:accuracy} Accuracy}

\begin{figure}
    \includegraphics[width=8.6 cm]{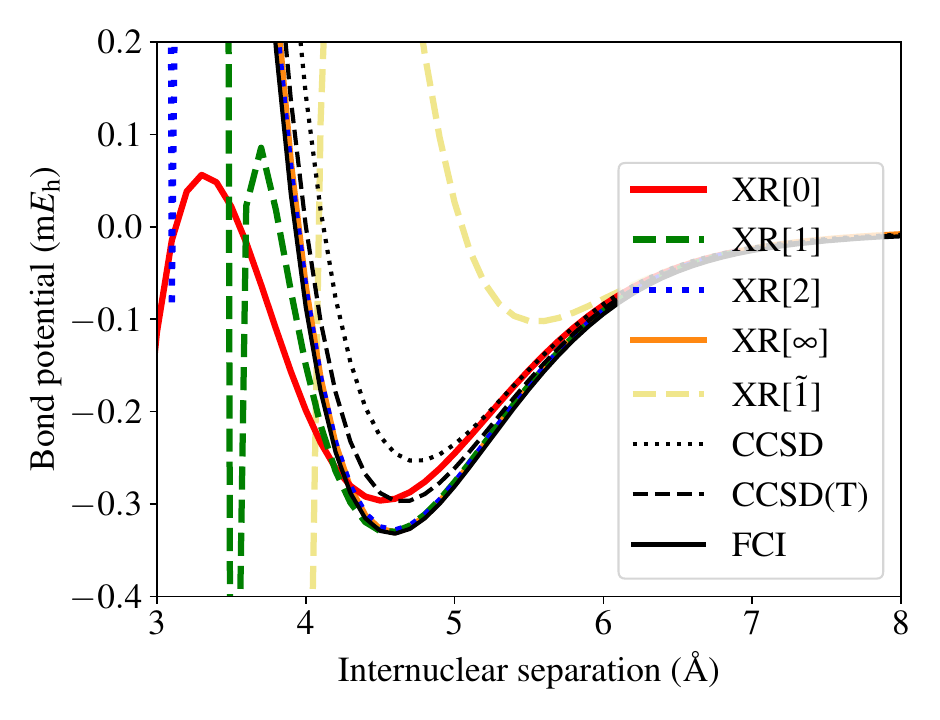}
    \caption{\label{fig:XR0vsXR1}The dissociation curve of the 6-31G Be dimer is plotted for variants of the XR method, differing in the order to which the $\hat{S}$ operator is computed.  Marked improvement is shown
     with each
     order, and the importance of using a consistent order in $\boldsymbol{\sigma}$ is demonstrated by comparison to XR[$\tilde{1}$].
    The XR methods all use the same underlying set of 23 many-electron states per Be atom.  Standard electronic structure results are provided for reference.}
\end{figure}

\begin{figure}
    \centering
    \includegraphics[width=8.6 cm]{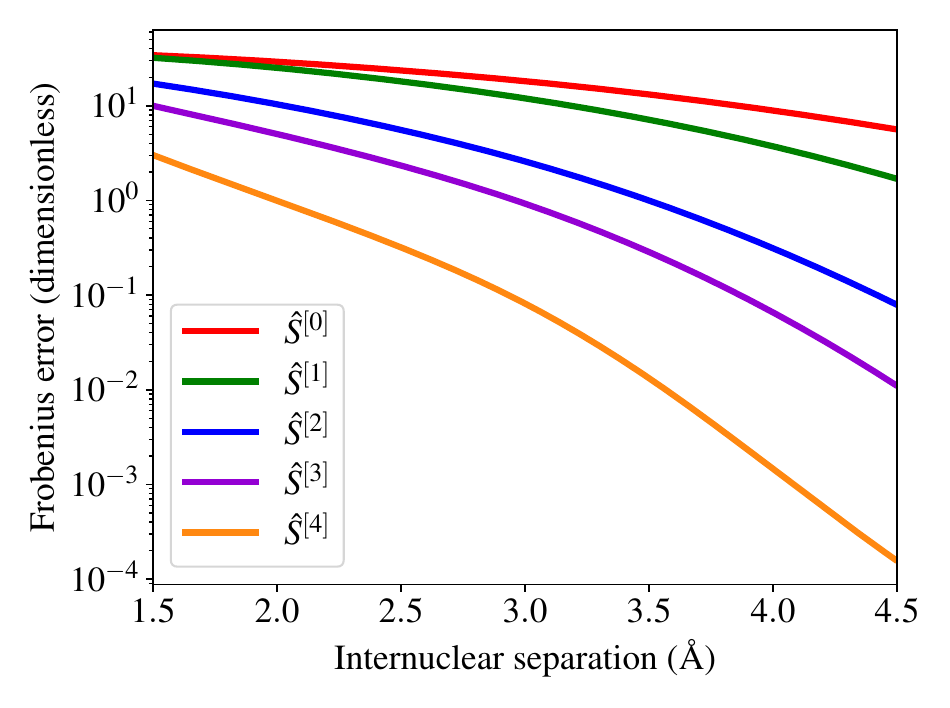}
    \caption{\label{fig:s_convergence}For the same system as in Fig.~\ref{fig:XR0vsXR1}, the Frobenius norm of the deviation of the $(23 \cdot 23) \times (23 \cdot 23)$ many-body overlap matrix $\mathbf{S}$ is shown for
    expansion of $\hat{S}$ up to increasingly higher orders $\hat{S}^{[o]}$.
    The convergence of the series is seen to be very quick, as expected from its exponential structure.  The jumps in accuracy with even orders can be rationalized in terms of properly accounting for Pauli repulsions.}
\end{figure}

To test the numerical accuracy of the scheme proposed here, we choose a dimer of 6-31G Be atoms, using exactly the same set of 23 many-electron states per atom as described in previous work\cite{Liu.2019} (11 neutral, 4 cationic, 8 anionic).
The excitonic Hamiltonian is completely described by the combination of 
 these states (parallel to defining the single-electron basis)
 with the system geometry 
 and the XR method variant.
For all calculations in this work, the ground state of such a Hamiltonian is
 found via the previously described excitonic CCSD (exact for our dimer Hamiltonians), giving in full: XR2[$o$]-CCSD,
 for varied order $o$.
Suppressing the nonvariable method descriptors, 
 we will show results for XR[0] (numerically identical to XR for our cases),
 XR[1], XR[2], and XR[$\infty$] (implemented as previously described for XR$^\prime$).

For comparison, we use the following standard electronic structure methods: conventional CCSD, also with perturbative triples [CCSD(T)], and FCI. These conventional coupled-cluster results, along with the fragment Hartree-Fock (HF) orbitals and supersystem molecular integrals were obtained from the Psi4 program package (versions 1.4 to 1.6).\cite{Smith.2020}
The FCI calculations were done with in-house code using the same integrals.

In Fig.~\ref{fig:XR0vsXR1}, the Be dimer dissociation curve is shown for each of the aforementioned XR[$o$] variants and conventional methods.  Due to the optimal choice of the 23 single-atom states, the XR[$\infty$] result is seen to be essentially exact (as has been shown previously), illustrating that the errors in
 XR[0] through XR[2]
 originate from the excitonic treatment, and not the states.

XR[0] yields an accuracy comparable to that of CCSD(T) for large distances down to the equilibrium distance, with CCSD(T) overestimating and XR[0] underestimating the equilibrium distance by roughly 0.07 \AA~and 0.09 \AA, respectively. 
Both methods underestimate the binding energy by about $3.7$$\times$$10^{-5}$ Hartree ($E_\text{h}$).
However, at shorter distances than the equilibrium distance, XR[0] highly underestimates the repulsive part of the curve, until it behaves qualitatively wrong at yet shorter distances. This is the same as what is obtained from an implementation of XR proper, which is being reported for the first time.  The unacceptability of this result is what has motivated the current work on XR[$o$] methods.

Since these results are new, it is worthwhile to give a short illustration of how the severity of this error arises.
Consider the simple model of a He atom that is given only its HF configuration as a model space.
Let this system be in the vicinity of any empty spatial orbital.
This can even belong to a so-called ghost atom, with no nucleus or electrons of its own.
Although the true energy of this system is still just the HF energy of He, the XR-Hamiltonian expectation value for this system (diagonal element in a non-Hermitian resolution) will be corrupted.
Due to the biorthogonalization, the complements of the original He molecular orbitals each contain a component of the new spatial orbital, with the same spin.
When the expectation value is taken, this effectively introduces off-diagonal-like scattering terms into the energy expression, which are not balanced by any consideration of energy gap; they
 enter
 with their raw coupling strengths.
These can be quite large, since they represent two tightly packed electrons attempting to escape each other.

Something like the forgoing will happen for every core pair that overlaps a mostly empty orbital of a neighboring fragment, and we have only illustrated one of the many ways that a projective energy expression can be corrupted.
This illustration has also pointed the way forward that we have taken:  the building of a hierarchy that tends towards a Hermitian effective Hamiltonian, meaning that the bras and kets of the representation span the same space (even if, as mentioned above, the \textit{representation} is not Hermitian, due to the non-orthonormal bases).

With XR[1], we move one step toward this Hermitian Hamiltonian that has demonstratively satisfactory numerics. 
The dissociation curve in Fig.~\ref{fig:XR0vsXR1} appears to be converging quickly to the FCI result.
XR[1] aligns very well with FCI around the equilibrium distance, where the energy error at the minimum reduces from XR[0] to XR[1] by roughly 95 \%.
Even at shorter distances XR[1] performs reasonably well inside the rise above the dissociation energy threshold, after which it is abruptly corrupted.
XR[2] is even more accurate, with the result being a faithful representation of the FCI potential at yet shorter bond distances.
It is anticipated from this that convergence to the exact result will be quick with respect to order.

An additional numerical experiment labeled XR[$\tilde{1}$] also appears in Fig.~\ref{fig:XR0vsXR1}.
This curve was produced by expanding $\hat{S}$ in eq.~\ref{eq:termwisehermitian} to first order,
which, as explained immediately following that equation, has the effect of mixing different orders in $\boldsymbol{\sigma}$.
The extent of the degradation of the results is not easily anticipated because XR[$\tilde{1}$] is essentially XR[1] with select higher-order terms added, providing an exactly Hermitian effective Hamiltonian already at
 (apparent)
 first order.
This result is, however, consistent with the degradation of results that have been observed when $\mathbf{S}$ (before inversion) is computed to higher order than $\tilde{\mathbf{H}}$ in eq.~\ref{eq:matrixelements}.
Simply put, internal inconsistencies with respect to the order of $\hat{S}$ are not tolerated.

With respect to the nature of the sudden breakdown
 of all XR methods at short bond distances,
 this may originate from the system under investigation, rather than represent a general behavior.
The Be dimer is known to be a tricky problem\cite{ElKhatib.2014, Sharma.2014} that is only handled well by high levels of theory with at least a triple-zeta basis.
In fact, we have chosen this problem in part because it is a stringent test of method quality, having strong correlations within the atoms, in spite of its advantageously small number of electrons.
The experimental equilibrium distance is roughly two Angstroms shorter than even the FCI result in the present basis.
The exact 6-31G basis result appears to capture only a shoulder interaction of a well that is ten times deeper in reality, centered near 2.5 \AA.
With intermediate sized basis sets, these inner and outer interactions are even  separated by a barrier into distinct wells.
It could be that the interaction responsible for the inner well is being partially captured by the XR and XR[$o$] methods, being quite different from conventional methods, 
and this artifact degrades the stability of XR[$o$] in this region for small $o$.

Finally, in lieu of higher order XR[$o$] results, which will require a more efficient implementation than presently available, we study directly the convergence of the matrix $\mathbf{S}$ of overlaps for the model states of the dimer.
Since each atom has 23 possible states, the edge dimension of this matrix is $23 \cdot 23$.
Along with our brute-force implementation of XR[$\infty$], we also have the exact $\mathbf{S}$ available, and so Fig.~\ref{fig:s_convergence} shows the Frobenius norm of the deviation from exactness as a function of bond distance and order of approximation for $\hat{S}$.
The series is
 again
 seen to converge quickly with order, as we have
 also
 anticipated due to the relationship of the series for $\hat{S}$ to an exponential of an operator $\hat{\sigma}$ whose matrix elements are all less than 1 (often much less).

The jumps in accuracy of $\mathbf{S}$ with even orders have a straightforward rationalization.
At first order, for example, the single powers of interfragment orbital overlaps contribute to off-diagonal elements of $\mathbf{S}$ where the bra is in a charge-transfer state relative to the ket.
But they do not account for the fact that overlapping two occupied orbitals causes a reduction in the norm of an antisymmetrized product state.
This is present in diagonal terms that conserve charges of individual fragments and require two powers of interfragment overlaps to do so.
Attaching to an associated concept, the increase in energy that results from electrons mutually excluding each other is called Pauli repulsion.
This is consistent with the dramatically better accuracy of XR[2] at shorter bond distances, relative to XR[1], which is effectively missing Pauli repulsions.

\subsection{\label{sec:scaling} Computational Considerations}

Starting from XR[0], which scales with the fourth power of the fragment one-electron basis size for both the the memory usage and arithmetic, these scalings each increase by one power with each additional order in the approximation of $\hat{S}$.
These scalings can be straightforwardly extracted from the term containing the largest transition density-tensor for a given XR[$o$].
Starting with eq.~\ref{eq:v1111} for XR[0] (a component of
 eq.~\ref{eq:S0Hdimer}), 
each subsequent order yields one additional index on each density tensor (never two, because diagonal blocks of $\boldsymbol{\sigma}$ are zero), giving eqs.~\ref{eq:s12_v1111}--\ref{eq:s12_v2222} (in eq.~\ref{eq:S1Hdimer}) for XR[1],
 eqs.~\ref{eq:s12_s21_v1111} \& \ref{eq:s12_s12_v1111}--\ref{eq:s12_s12_v2222} (in eq.~\ref{eq:S2Hdimer}) for XR[2],
 and so on, increasing by one the dimensionality of the largest transition-density tensor to be stored.
Since no new contraction types are introduced (just an increasing number of contractions with the two-index $\boldsymbol{\sigma}$), this also increments arithmetic scaling by one.
Note
 that these scalings do not include the number of states used for the calculation, since the size of the truncated state spaces required to ensure a certain accuracy does not necessarily increase with the system size. 

XR[0] and XR[1] are therefore seen to have respective scalings that are better than standard CCSD(T) by three and two orders in their arithmetic complexity, as well as two and one orders in their memory usage.
XR[2] is one order faster than CCSD(T) with comparably scaling storage, achieving
 nearly FCI accuracy for bond distances around the equilibrium point and larger. 

Under the assumption that satisfactory states to describe interfragment interactions can generally be obtained (addressed in Sec.~\ref{sec:conclusions}), the most pressing concern about the structure of XR[$o$] theory itself is the high-dimensional transition-density tensors that will be required for those fragments.
Already for
 XR[2]
 with three charges allowed per fragment (cationic, neutral, and anionic),
 six-index
 tensors are required.
However, we presume that the amount of information in high-dimensional tensors that is not fundamentally redundant with information in low-dimensional tensors is ever smaller for the higher-dimensional tensors.  In the limit of single-determinant states, there is no unique information beyond the one-electron reduced density matrices for each state.
To the extent that high-dimensional tensors do contain unique information, it is anyway folly to consider throwing them away out of convenience, rather than attempting to isolate and compactly express the interesting part.
A more complete set of numerical experiments involving compression schemes and cumulant-like expansions is necessary, but, as a quick demonstration of what is possible, we apply a simple scheme using singular-value decomposition, which is so effective as to have become the default mode in our implementation.

First, a matricization of each density tensor separates the creation indices from the annihilation indices into the respective two matrix axes.
This matrix is then internally represented as a contraction of two matrices, each consisting of only the respective left or right singular vectors of the original matrix that correspond to singular values above $10^{-6}$, with each vector rescaled by the square root of its associated singular value.
This product
 matrix is then properly reindexed as the higher-dimensional tensor.
This fast-and-loose approach already saves a factor of $\sim$$20$ in storage requirements, with only a $10^{-12}~E_\text{h}$ deviation in the XR[1] energy at 4.5 \AA, for
 example.
 XR[2] without the compression scheme was not even attempted.
Clearly we could tolerate a yet looser threshold, but more thoughtful schemes are anyway needed.
This simply demonstrates the extremely low information density of these tensors, making our forgoing arguments quite plausible.
Ultimately, we will seek to implicitly represent approximated high-dimensional tensors, which have never been computed, in terms of linear combinations of antisymmetrized outer products of the low-dimensional density tensors that are explicitly known, which are themselves stored in a compressed representation. Since this is necessary for other methods as well, this field has been subject of extensive research for many decades, and has already yielded promising results.\cite{Mazziotti.1999, Mazziotti.1999b}

\section{\label{sec:conclusions}Conclusions and Outlook}

Building on earlier proof-of-concept work regarding the applicability and efficiency of using excitonically renormalized Hamiltonians, this article has focused specifically on the building of the Hamiltonian.
A previously \textit{ad hoc} approximation for this that outperformed its parent theory, but which could only work for toy systems, has been placed on a solid theoretical footing as the limit of a systematically improvable hierarchy of tractable approximations.

This has also been shown to numerically converge quickly.
For the system under consideration, the zeroth-order result already yields an accuracy comparable with that of CCSD(T) around the binding minimum, while the first
 and second order results yield
 almost FCI accuracy in this region and at larger
 distances, with the second-order result also being accurate quite high up the inner repulsive wall.
The scaling of the zeroth-order method is comparable to conventional HF, while, for the
 second-order
 method, the memory scaling increases by
 two powers,
 and is comparable to
 that of CCSD(T). 

If similar accuracy can be obtained for other systems, the newly presented methods will provide a tool for highly precise \textit{ab initio} calculations.
Due to the way that intrafragment complexity is modularized, it could prove extremely useful for applications where more than one fragment needs to be treated at a multi-reference level.
Such systems are often encountered in catalysis, especially for the activation of small molecules.\cite{Becker.2018, Wang.2023, Rosch.2021, Rang.2024, Piquemal.2006}
In principle this can also be provided by ASD and BCCC, but these are, as already mentioned, limited by the preliminary of a globally orthogonal orbital basis, which is required to be fully localized on the fragments. This requirement cannot be fulfilled in general, especially if many virtual orbitals in the active space are necessary for obtaining accurate results, since these can often not be localized well. Systems like these are regularly encountered in the targeted transition metal based applications. Furthermore, correlation outside of the active spaces is also included with XR right away, whereas BCCC and ASD only treat them on the mean-field level. Hence, XR can be seen as a generalization of ASD, with ASD being the special case of an active-space XR[0] with globally orthogonal and fragment-local state spaces.

The question naturally arises, whether these methods will perform similarly well for other systems, or if the low-order variants here only reproduce the bonding well so accurately because the bond distance is very large and/or because the systems are small.
Exploring these questions will require an improved implementation, which is already in progress and will be the subject of future publications.

Of the two most prominent outstanding issues, one is the improvement of the XR[$o$] Hamiltonian build itself, developing sound approximations for high-dimensional transition density tensors, and compression schemes for the explicit tensors. Second, assuming modest to good success of this methodology, obtaining good states for fragments will be an enduring area of work; that is, we require a small number of fragment states that adequately capture their interactions with neighbors.  In addition to simply obtaining these states as a permanent resource for future supersystem Hamiltonian builds, theoretical work needs to first be done on the best ways to obtain them.

\section*{Acknowledgments}

ADD would like to thank Brian Austin, Oriol Vendrell, Markus Schröder, Sudip Sasmal, Dominic Kempf, Christoph Bannwarth, Luka Dockx, and Christopher D. Goff for helpful conversations.  MB and AD acknowledge support through the collaborative research center CRC 1249 ``\textit{N}-heteropolycycles as functional materials'' of the German Science Foundation.  ADD acknowledges support from the Hornage Fund at the University of the Pacific, as well as equipment and travel support provided by the Dean of the College of the Pacific.

\section*{Data Availability Statement}

The data that support the findings of this study are available from the corresponding author upon reasonable request.

\appendix

\section{\label{sec:notational_diffs} Notational Deviations from Previous Articles}

This article has some slight notational changes from previous publications.\cite{Dutoi.2019, Liu.2019}
The Hamiltonian in eq.~\ref{eq:excitonic} has been modified from its presentation in the first article, where $\hat{H_1}$, $\hat{H_2}$, $\hat{H_3}$, and $\hat{H_4}$ were used in their entirety in the respective matrix elements for the monomer, dimer, trimer, and tetramer terms, relying on vanishing actions of field operators to implicitly restrict summations.
This change has been made so that, for example, the effective monomer Hamiltonian for fragment $m$ does not incorporate the attraction to nuclei of other fragments; this attraction now appears among the dimer terms.
As a philosophical note, the original notation could be recovered by treating the nuclei (in delta-position states) on the same footing as the electrons.
The original article also used $\hat{H}^{(m_1,m_2)}$, instead of the $\hat{H}_{\{m_1,m_2\}}$, \textit{etc.} presently defined in eqs.~\ref{eq:abinitiofragdecomp1} and \ref{eq:abinitiofragdecomp2}, which was not as clear about the symmetry of these terms under fragment permutation, and not consistent in the use of subscripts for qualification of Hamiltonian components.

Tangentially, concerning the nuclei,
it is also sensible to add the scalar energies from intrafragment nuclear repulsion to the monomer terms and the interfragment nuclear repulsion to the dimer terms, but following this decision through the derivations for all method variants will verify the familiar case that the result will not be different if the entire nuclear repulsion is accounted for as an addendum to the electronic structure calculation.

Finally, regarding the nomenclature, since it was noted that the \textit{ad hoc}  XR$^\prime$ approximation corresponds precisely to XR[$\infty$] in the present scheme,  
the abbreviation XR$^\prime$ is henceforth superseded and retired.

\section{\label{sec:diagrams}Working Equations for Diagrams and Matrix Elements}

\subsection{\label{sec:work_eq_ex}Diagram Evaluation}

It will be instructive to explicitly derive one of the matrix elements of the form of eqs.~\ref{eq:explicitmatrixelementA} and \ref{eq:explicitmatrixelementB}.
Among the possible choices, the dimer element of $\hat{S}^{[3]}$ most compactly exhibits all of the features we would like to discuss.
We use the convention that repeated indices are contracted, irrespective of whether they are sub- or super-scripts.
Choosing fragments 1 and 2, we obtain for this
\begin{eqnarray}
    \label{eq:deriveelementA}
    &~& \langle\psi^{i_{1}} \psi^{i_{2}}|
    \sigma_{p q} \sigma_{r s} \sigma_{t u} \hat{c}_{p} \hat{c}_{r} \hat{c}_{t} \hat{a}^{u} \hat{a}^{s} \hat{a}^{q}
    |\psi_{j_{1}} \psi_{j_{2}}\rangle
    \\
    &~& \nonumber \\
    \label{eq:deriveelementB}
    =
    &~& \sigma_{p_1 q_2} \sigma_{r_1 s_2} \sigma_{t_1 u_2}~{}^{i_{1} i_{2}}\langle\hat{c}_{p_1} \hat{c}_{r_1} \hat{c}_{t_1} \hat{a}^{u_2} \hat{a}^{s_2} \hat{a}^{q_2}\rangle_{j_{1} j_{2}}
    \\
    &-& \sigma_{p_1 q_2} \sigma_{r_1 s_2} \sigma_{t_2 u_1}~{}^{i_{1} i_{2}}\langle\hat{c}_{p_1} \hat{c}_{r_1} \hat{a}^{u_1} \hat{c}_{t_2} \hat{a}^{s_2} \hat{a}^{q_2}\rangle_{j_{1} j_{2}} \nonumber
    \\
    &-& \sigma_{p_1 q_2} \sigma_{r_2 s_1} \sigma_{t_1 u_2}~{}^{i_{1} i_{2}}\langle\hat{c}_{p_1} \hat{c}_{t_1} \hat{a}^{s_1} \hat{c}_{r_2} \hat{a}^{u_2} \hat{a}^{q_2}\rangle_{j_{1} j_{2}} \nonumber
    \\
    &+& \sigma_{p_1 q_2} \sigma_{r_2 s_1} \sigma_{t_2 u_1}~{}^{i_{1} i_{2}}\langle\hat{c}_{p_1} \hat{a}^{u_1} \hat{a}^{s_1} \hat{c}_{r_2} \hat{c}_{t_2} \hat{a}^{q_2}\rangle_{j_{1} j_{2}} \nonumber
    \\
    &-& \sigma_{p_2 q_1} \sigma_{r_1 s_2} \sigma_{t_1 u_2}~{}^{i_{1} i_{2}}\langle\hat{c}_{r_1} \hat{c}_{t_1} \hat{a}^{q_1} \hat{c}_{p_2} \hat{a}^{u_2} \hat{a}^{s_2}\rangle_{j_{1} j_{2}} \nonumber
    \\
    &+& \sigma_{p_2 q_1} \sigma_{r_1 s_2} \sigma_{t_2 u_1}~{}^{i_{1} i_{2}}\langle\hat{c}_{r_1} \hat{a}^{u_1} \hat{a}^{q_1} \hat{c}_{p_2} \hat{c}_{t_2} \hat{a}^{s_2}\rangle_{j_{1} j_{2}} \nonumber
    \\
    &+& \sigma_{p_2 q_1} \sigma_{r_2 s_1} \sigma_{t_1 u_2}~{}^{i_{1} i_{2}}\langle\hat{c}_{t_1} \hat{a}^{s_1} \hat{a}^{q_1} \hat{c}_{p_2} \hat{c}_{r_2} \hat{a}^{u_2}\rangle_{j_{1} j_{2}} \nonumber
    \\
    &-& \sigma_{p_2 q_1} \sigma_{r_2 s_1} \sigma_{t_2 u_1}~{}^{i_{1} i_{2}}\langle\hat{a}^{u_1} \hat{a}^{s_1} \hat{a}^{q_1} \hat{c}_{p_2} \hat{c}_{r_2} \hat{c}_{t_2}\rangle_{j_{1} j_{2}} \nonumber
    \\
    &~& \nonumber \\
    \label{eq:deriveelementC}
    =
    &~& (-1)^{n_{j_1}} ~ \sigma_{p_1 q_2} \sigma_{r_1 s_2} \sigma_{t_1 u_2} ~ \rho_{p_1 r_1 t_1} \, \rho^{u_2 s_2 q_2}
    \\
    &-& (-1)^{n_{j_1}} ~ \sigma_{p_1 q_2} \sigma_{r_1 s_2} \sigma_{t_2 u_1} ~ \rho_{p_1 r_1}^{u_1} \, \rho_{t_2}^{s_2 q_2} \nonumber
    \\
    &-& (-1)^{n_{j_1}} ~ \sigma_{p_1 q_2} \sigma_{r_2 s_1} \sigma_{t_1 u_2} ~ \rho_{p_1 t_1}^{s_1} \, \rho_{r_2}^{u_2 q_2} \nonumber
    \\
    &+& (-1)^{n_{j_1}} ~ \sigma_{p_1 q_2} \sigma_{r_2 s_1} \sigma_{t_2 u_1} ~ \rho_{p_1}^{u_1 s_1} \, \rho_{r_2 t_2}^{q_2} \nonumber
    \\
    &-& (-1)^{n_{j_1}} ~ \sigma_{p_2 q_1} \sigma_{r_1 s_2} \sigma_{t_1 u_2} ~ \rho_{r_1 t_1}^{q_1} \, \rho_{p_2}^{u_2 s_2} \nonumber
    \\
    &+& (-1)^{n_{j_1}} ~ \sigma_{p_2 q_1} \sigma_{r_1 s_2} \sigma_{t_2 u_1} ~ \rho_{r_1}^{u_1 q_1} \, \rho_{p_2 t_2}^{s_2} \nonumber
    \\
    &+& (-1)^{n_{j_1}} ~ \sigma_{p_2 q_1} \sigma_{r_2 s_1} \sigma_{t_1 u_2} ~ \rho_{t_1}^{s_1 q_1} \, \rho_{p_2 r_2}^{u_2} \nonumber
    \\
    &-& (-1)^{n_{j_1}} ~ \sigma_{p_2 q_1} \sigma_{r_2 s_1} \sigma_{t_2 u_1} ~ \rho^{u_1 s_1 q_1} \, \rho_{p_2 r_2 t_2}   \nonumber
    \\
    &~& \nonumber \\
    \label{eq:deriveelementD}
    =
    &~& ~\,(-1)^{n_{j_1}}~\rho_{p_1 r_1 t_1}^{} \, \rho_{}^{u_2 s_2 q_2} ~ \sigma_{p_1 q_2} \sigma_{r_1 s_2} \sigma_{t_1 u_2}
    \\
    &-&   3(-1)^{n_{j_1}}~\rho_{p_1 r_1}^{u_1} \, \rho_{t_2}^{s_2 q_2} ~ \sigma_{p_1 q_2} \sigma_{r_1 s_2} \sigma_{t_2 u_1} \nonumber
    \\
    &+&   3(-1)^{n_{j_1}}~\rho_{p_2 r_2}^{u_2} \, \rho_{t_1}^{s_1 q_1} ~ \sigma_{p_2 q_1} \sigma_{r_2 s_1} \sigma_{t_1 u_2} \nonumber
    \\
    &-& ~\,(-1)^{n_{j_1}}~\rho_{p_2 r_2 t_2}^{} \, \rho_{}^{u_1 s_1 q_1} ~ \sigma_{p_2 q_1} \sigma_{r_2 s_1} \sigma_{t_2 u_1} \nonumber
    \\
    &~& \nonumber \\
    \label{eq:deriveelementE}
    =
    &~&
    3! \big[
    \langle \sigma_{1 2} \sigma_{1 2} \sigma_{1 2} \rangle ~+~ \langle \sigma_{1 2} \sigma_{1 2} \sigma_{2 1} \rangle
    \\
    &~& +~ \langle \sigma_{2 1} \sigma_{2 1} \sigma_{1 2} \rangle ~+~ \langle \sigma_{2 1} \sigma_{2 1} \sigma_{2 1} \rangle 
    \big] \nonumber
\end{eqnarray}
The factorial in the last line cancels the inclusion of such a denominator in the eventual diagram definitions.
It is more intuitively thought of as living on the other sides of this equation, but is moved to declutter the progression.
We now go through this derivation line by line to highlight some of the details.

In moving from eqs.~\ref{eq:deriveelementA} to \ref{eq:deriveelementB}, we have broken the full range of summation for each index into subcases that restrict orbitals to specific fragments.
Operators acting on fragments other than 1 or 2 yield zero,
and we recall that diagonal blocks of $\boldsymbol{\sigma}$ are zero, shrinking the number of relevant cases to eight.
The indices outside the angle brackets in eq.~\ref{eq:deriveelementB} distinguish matrix elements in an abbreviated form from the notation for vacuum expectation values.
Finally, we have reordered the field operators to collect those for each fragment together, but without swapping those that act on the same fragment, accumulating the respective phase.

In moving from eqs.~\ref{eq:deriveelementB} to \ref{eq:deriveelementC}, we have written each of the dimer density-tensor elements as a product of monomer factors, with implicit state indices suppressed.
The procedure for doing this is exemplified for the first term via
\begin{eqnarray}
    \label{eq:howtofactorizeA}
    &~& {}^{i_{1} i_{2}}\langle\hat{c}_{p_1} \hat{c}_{r_1} \hat{c}_{t_1} \hat{a}^{u_2} \hat{a}^{s_2} \hat{a}^{q_2}\rangle_{j_{1} j_{2}}
    \\
    \label{eq:howtofactorizeB}
    &=& \langle \hat{\psi}^{i_2} \hat{\psi}^{i_1} \hat{c}_{p_1} \hat{c}_{r_1} \hat{c}_{t_1} \hat{a}^{u_2} \hat{a}^{s_2} \hat{a}^{q_2} \hat{\psi}_{j_1} \hat{\psi}_{j_2}\rangle 
    \\
    \label{eq:howtofactorizeC}
    &=& (-1)^{3 n_{j_1}}\langle \hat{\psi}^{i_2} \hat{\psi}^{i_1} \hat{c}_{p_1} \hat{c}_{r_1} \hat{c}_{t_1} \hat{\psi}_{j_1}  \hat{a}^{u_2} \hat{a}^{s_2} \hat{a}^{q_2}  \hat{\psi}_{j_2}\rangle 
    \\
    \label{eq:howtofactorizeD}
    &=& (-1)^{n_{j_1}}\langle \hat{\psi}^{i_1} \hat{c}_{p_1} \hat{c}_{r_1} \hat{c}_{t_1} \hat{\psi}_{j_1} \hat{\psi}^{i_2}  \hat{a}^{u_2} \hat{a}^{s_2} \hat{a}^{q_2} \hat{\psi}_{j_2} \rangle 
    \\
    \label{eq:howtofactorizeE}
    &=& (-1)^{n_{j_1}} ~ {}^{i_{1}}\langle\hat{c}_{p_1} \hat{c}_{r_1} \hat{c}_{t_1} \rangle_{j_1} {}^{i_{2}}\langle   \hat{a}^{u_2} \hat{a}^{s_2} \hat{a}^{q_2}  \rangle_{j_2} 
\end{eqnarray}
where, in order to to express this quantity as a vacuum expectation value in eq.~\ref{eq:howtofactorizeB}, we have used 
\begin{eqnarray}
    \label{eq:genstatecreation}
    \hat{\psi}_{i_1} &=& \sum_{P_1} z_{i_1}^{P_1} \hat{\phi}_{P_1} \\
    \label{eq:genstateannihilation}
    \hat{\psi}^{i_1} &=& \sum_{P_1} z_{i_1}^{{P_1}*} \hat{\phi}^{P_1} 
\end{eqnarray}
as defined in prior work, but for a presumedly Hermitian $\mathbf{z}$.
In eq.~\ref{eq:howtofactorizeC}, the operators for fragment 1 are now in a contiguous string, requiring $3n_{j_1}$ permutations of anticommuting field operators;
here we need only to invoke the assumption that the $|\psi_{i_m}\rangle$ have well-defined electron number.
Each term in each of the complete strings for fragment 1 (upon insertion of eqs.~\ref{eq:genstatecreation} and \ref{eq:genstateannihilation}) must contain an even number of operators to yield a nonzero result, and so we have the operators for fragment 2 collected in eq.~\ref{eq:howtofactorizeD} with no further accumulation of phase.
If the complete strings for fragments 1 and 2 are now each rewritten as the sum of a scalar and normal-ordered strings, which again requires insertion of eqs.~\ref{eq:genstatecreation} and \ref{eq:genstateannihilation}, then only the products of the scalar terms will survive the vacuum expectation value.
After recontraction of these expectation values the coefficients of eqs.~\ref{eq:genstatecreation} and \ref{eq:genstateannihilation}, these are the same as the isolated single-fragment matrix elements given in eq.~\ref{eq:howtofactorizeE}.
We recognize these matrix elements as the explicit forms of transition-density-tensor elements, 
completing our arrival at the first term of eq.~\ref{eq:deriveelementC}.

It is worth noting that the $(-1)^{n_{j_1}}$ phase factor, which only appears for matrix elements that transfer an odd number of electrons between fragments, is subject to a convention that is dependent on (\textit{i}) the relationship between the ket labels and the order in which the operators $\hat{\psi}_{j_1}$ and $\hat{\psi}_{j_1}$ are applied and (\textit{ii}) the decision to shuffle the central field operators into ascending or descending order by fragment.
The former choice actually affects the definitions of the states, flipping the signs of states where both fragments have odd numbers of electrons, and therefore, the signs of these matrix elements for them, which must be with another state having even numbers of electrons for both fragments.
If we change only this convention, we instead obtain a phase of $(-1)^{n_{i_2}}$ using similar logic to
 the progression in eqs.~\ref{eq:howtofactorizeA}--\ref{eq:howtofactorizeE};
this is indeed the opposite of $(-1)^{n_{j_1}}$ for a state with two odd numbers of electrons coupling to a state with two even numbers of electrons.
Conversely, if we change the sorting of the central field operators,
 using descending order by fragment, it should not change the overall result, and we obtain either $-(-1)^{n_{i_1}}$ or $-(-1)^{n_{j_2}}$, depending on whether we use the state-ordering convention of this work, or the alternative, respectively;
the minus signs account for swapping the two strings with odd numbers of operators.
As expected we have $(-1)^{n_{j_1}}$$=$$-(-1)^{n_{i_1}}$ and $(-1)^{n_{i_2}}$$=$$-(-1)^{n_{j_2}}$ for the couplings of odd-odd states with even-even states.
Finally, for odd total numbers of electrons (odd-even with even-odd), all conventions give the same phase, consistent with the fact that neither the bra nor ket changes sign dependent on any convention.
In summary, in spite of these equations apparently privileging one of the states in either the bra or ket, the symmetry lies in the fact that any one of these numbers of electrons may be chosen, as long as it is used consistently, and some choices are completely equivalent for all nonzero elements.

To get from eqs.~\ref{eq:deriveelementC} to \ref{eq:deriveelementD}, we recognize some terms as redundant simply by swapping the letters used for dummy indices, but without swapping fragment labels.
We have furthermore reordered the scalar factors in these terms to emphasize the character of the density tensors.
For example, the first and fourth terms will only be nonzero for triple-charge-transfer fluctuations, whereas the second and third terms are nonzero for single charge transfers.
It is also apparent that these terms are related by permutations of fragments 1 and 2, which changes the direction of charge transfer.
However, this permutation comes with a sign change that cannot itself be written neatly as a function of the permutation [\textit{i.e.}, $(-1)^{n_{j_1}}$$\neq$$-(-1)^{n_{j_2}}$] because, in spite of the other symmetries between fragments 1 and 2 in the starting point, this phase is anchored to a state-ordering convention that is globally fixed.
Retaining the $(-1)^{n_{j_1}}$ convention, this sign change can be traced to the additional odd number of permutations necessary to shuffle the field operators into ascending order when the direction of charge transfer changes.
One could choose the numerically identical descending-order convention for the field operators, but, due to the fixed state-ordering convention, it will nevertheless result in a phase expressed in terms of fragment 1 as $(-1)^{n_{i_1}}$.

In eq.~\ref{eq:deriveelementE}, we finally express, in the same order, each of the terms from eq.~\ref{eq:deriveelementD} using the abbreviated notation for diagrams.
The diagrams include any relevant phases and redundancy factors, as well as the factorial denominator from the exponential-like expansion of $\hat{S}$.
As just noted, the working equation to algebraically resolve any diagram is identical, to within a phase, to the expression for the same diagram with fragment labels permuted.
Therefore, we will not record the working equations for diagrams that can be generated as fragment permutations of other recorded diagrams.
A key deduction from the above, for dimer matrix elements, is that electron-number-dependent phases will appear for any elements that represent the transfer of an odd number of electrons between fragments, and only such elements, and there will always be an additional sign change in the formulas for these when the fragments are permuted, with the original phase expression not having its fragment label exchanged.
For dimers, this rule then suffices to reproduce formulas for the missing permuted diagrams to come.
The rules for trimer diagrams and beyond are more complex and will be addressed in a future publication that harnesses the power of a truly diagrammatic representation.

\subsection{\label{sec:work_eqs}Working Equations}

Listed here are all of the diagrams needed to produce the results in this work.
Pure overlap diagrams up to fourth order in $\boldsymbol{\sigma}$ are given, as are all
 first- and second-order
 diagrams containing Hamiltonian integrals.
Both of these listings include some diagrams that were unnecessary to implement for this article because our neutral, cationic, and anionic states differ from each other by a maximum of two electrons.
The complete working equations for XR, XR[0],
 XR[1], and XR[2]
 matrix elements are then given in terms of these diagrams. 

We do not separately give diagrams for $\hat{t}$ and ${}^\alpha\hat{u}$ or ${}^\mathbf{m}\hat{u}$, as these are trivially related to the generic diagrams for one-electron operators given for $\hat{h}$.
For completeness of the system, we list the sole zero-fragment diagram as well, whose abbreviation is consistent with its evaluation as the vacuum expectation value of no operators.
This diagram would formally occur in the leading order term of the exponential-like expansion for $\hat{S}$, where it will eventually be multiplied by Kronecker deltas for the fragment states.
\begin{eqnarray}
    \langle \rangle &=& 1
    \\
    \langle \sigma_{12} \rangle &=&
    (-1)^{n_{j_1}}
    \rho_{p_1} \, \rho^{q_2} \, \sigma_{p_1 q_2}  
    \\
    \langle \sigma_{12} \sigma_{21} \rangle &=& 
    -\rho_{p_1}^{s_1} \, \rho_{r_2}^{q_2} \, \sigma_{p_1 q_2}  \sigma_{r_2 s_1}  
    \\
    \langle \sigma_{12} \sigma_{12} \rangle &=&
    \frac{1}{2}
    \rho_{p_1 r_1} \, \rho^{s_2 q_2} \, \sigma_{p_1 q_2}  \sigma_{r_1 s_2}  
    \\
    \langle \sigma_{12} \sigma_{12} \sigma_{21} \rangle &=&
    \\ &~&
    -\frac{(-1)^{n_{j_1}}}{2}
    \rho_{p_1 r_1}^{u_1} \, \rho_{t_2}^{s_2 q_2} \, \sigma_{p_1 q_2}  \sigma_{r_1 s_2}  \sigma_{t_2 u_1} \nonumber
    \\
    \langle \sigma_{12} \sigma_{12} \sigma_{12} \rangle &=&
    \\ &~&
    \frac{(-1)^{n_{j_1}}}{6}
    \rho_{p_1 r_1 t_1} \, \rho^{u_2 s_2 q_2} \, \sigma_{p_1 q_2} \sigma_{r_1 s_2} \sigma_{t_1 u_2} \nonumber
    \\
    \langle \sigma_{12} \sigma_{12} \sigma_{21} \sigma_{21} \rangle &=& 
    \\ &~&
    \frac{1}{4}
    \rho_{p_1 r_1}^{w_1 u_1} \, \rho_{t_2 v_2}^{s_2 q_2} \, \sigma_{p_1 q_2}  \sigma_{r_1 s_2}  \sigma_{t_2 u_1} \sigma_{v_2 w_1} \nonumber
    \\
    \langle \sigma_{12} \sigma_{12} \sigma_{12} \sigma_{21} \rangle &=& 
    \\ &~&
    -\frac{1}{6}
    \rho_{p_1 r_1 t_1}^{w_1} \, \rho_{v_2}^{u_2 s_2 q_2} \, \sigma_{p_1 q_2}  \sigma_{r_1 s_2}  \sigma_{t_1 u_2}  \sigma_{v_2 w_1} \nonumber
    \\
    \langle \sigma_{12} \sigma_{12} \sigma_{12} \sigma_{12} \rangle &=&
    \\ &~&
    \frac{1}{24}
    \rho_{p_1 r_1 t_1 v_1} \, \rho^{w_2 u_2 s_2 q_2} \, \sigma_{p_1 q_2} \sigma_{r_1 s_2} \sigma_{t_1 u_2} \sigma_{v_1 w_2} \nonumber
\end{eqnarray}

\begin{eqnarray}
    \langle h^{1}_{1} \rangle &=& 
    \rho_{p_1}^{q_1} \, h^{p_1}_{q_1}
    \\
    \langle h^{1}_{2} \rangle &=&  
    (-1)^{n_{j_1}}  \rho_{p_1} \, \rho^{q_2} \, h^{p_1}_{q_2}
    \\
    \langle \sigma_{12} h^{2}_{1} \rangle &=&  
    -\rho_{t_1}^{q_1} \, \rho_{p_2}^{u_2} \, \sigma_{t_1 u_2}  h^{p_2}_{q_1}
    \\
    \langle \sigma_{12} h^{1}_{1} \rangle &=&  
    -(-1)^{n_{j_1}}  \rho_{p_1 t_1}^{q_1} \, \rho^{u_2} \, \sigma_{t_1 u_2}  h^{p_1}_{q_1}
    \\
    \langle \sigma_{12} h^{2}_{2} \rangle &=&  
    -(-1)^{n_{j_1}}  \rho_{t_1} \, \rho_{p_2}^{u_2 q_2} \, \sigma_{t_1 u_2}  h^{p_2}_{q_2}
    \\
    \langle \sigma_{12} h^{1}_{2} \rangle &=&
    \rho_{p_1 t_1} \, \rho^{u_2 q_2} \, \sigma_{t_1 u_2}  h^{p_1}_{q_2}  
    \\
    \langle \sigma_{1 2} \sigma_{2 1} h^{1}_{1} \rangle &=&
    -\rho_{p_1 t_1}^{w_1 q_1} \, \rho_{v_2}^{u_2} \, \sigma_{t_1 u_2} \, \sigma_{v_2 w_1} \, h^{p_1}_{q_1}
    \\
    \langle \sigma_{1 2} \sigma_{2 1} h^{1}_{2} \rangle &=&
    \\ &~&
    -(-1)^{n_{j_1}} \rho_{p_1 t_1}^{w_1} \, \rho_{v_2}^{u_2 q_2} \, \sigma_{t_1 u_2} \, \sigma_{v_2 w_1} \, h^{p_1}_{q_2} \nonumber
    \\
    \langle \sigma_{1 2} \sigma_{1 2} h^{2}_{1} \rangle &=&
    \\ &~&
    -\frac{(-1)^{n_{j_1}}}{2} \rho_{t_1 v_1}^{q_1} \, \rho_{p_2}^{w_2 u_2} \, \sigma_{t_1 u_2} \, \sigma_{v_1 w_2} \, h^{p_2}_{q_1} \nonumber
    \\
    \langle \sigma_{1 2} \sigma_{1 2} h^{1}_{1} \rangle &=&
    \frac{1}{2}\rho_{p_1 t_1 v_1}^{q_1} \, \rho_{}^{w_2 u_2} \, \sigma_{t_1 u_2} \, \sigma_{v_1 w_2} \, h^{p_1}_{q_1}
    \\
    \langle \sigma_{1 2} \sigma_{1 2} h^{2}_{2} \rangle &=&
    \frac{1}{2}\rho_{t_1 v_1} \, \rho_{p_2}^{w_2 u_2 q_2} \, \sigma_{t_1 u_2} \, \sigma_{v_1 w_2} \, h^{p_2}_{q_2}
    \\
    \langle \sigma_{1 2} \sigma_{1 2} h^{1}_{2} \rangle &=&
    \\ &~&
    \frac{(-1)^{n_{j_1}}}{2} \rho_{p_1 t_1 v_1}^{} \, \rho_{}^{w_2 u_2 q_2} \, \sigma_{t_1 u_2} \, \sigma_{v_1 w_2} \, h^{p_1}_{q_2} \nonumber
\end{eqnarray}

\begin{eqnarray}
    \label{eq:v1111}
    \langle v^{11}_{11} \rangle &=&  
    \rho_{p_1 q_1}^{s_1 r_1} \, v^{p_1 q_1}_{r_1 s_1}
    \\
    \langle v^{12}_{12} \rangle &=&
    4\, \rho_{p_1}^{r_1} \, \rho_{q_2}^{s_2} \, v^{p_1 q_2}_{r_1 s_2}
    \\
    \langle v^{11}_{21} \rangle &=&  
    2  (-1)^{n_{j_1}}  \rho_{p_1 q_1}^{s_1} \, \rho^{r_2} \, v^{p_1 q_1}_{r_2 s_1}
    \\
    \langle v^{12}_{11} \rangle &=&
    2(-1)^{n_{j_1}} \rho_{p_1}^{s_1 r_1} \, \rho_{q_2} \, v^{p_1 q_2}_{r_1 s_1}
    \\
    \langle v^{11}_{22} \rangle &=&  
    \rho_{p_1 q_1} \, \rho^{s_2 r_2} \, v^{p_1 q_1}_{r_2 s_2}
    \\
    \langle \sigma_{12} v^{12}_{11} \rangle &=&
    -2\, \rho_{p_1 t_1}^{s_1 r_1} \, \rho_{q_2}^{u_2} \, \sigma_{t_1 u_2} v^{p_1 q_2}_{r_1 s_1}
    \\
    \langle \sigma_{12} v^{22}_{21} \rangle &=&
    -2\, \rho_{t_1}^{s_1} \, \rho_{p_2 q_2}^{u_2 r_2} \, \sigma_{t_1 u_2} v^{p_2 q_2}_{r_2 s_1}
    \\
    \langle \sigma_{12} v^{12}_{12} \rangle &=&
    4 (-1)^{n_{j_1}} \rho_{p_1 t_1}^{r_1} \, \rho_{q_2}^{u_2 s_2} \, \sigma_{t_1 u_2} v^{p_1 q_2}_{r_1 s_2}
    \\
    \label{eq:s12_v1111}
    \langle \sigma_{12} v^{11}_{11} \rangle &=&  
    (-1)^{n_{j_1}}  \rho_{p_1 q_1 t_1}^{s_1 r_1} \, \rho^{u_2} \, \sigma_{t_1 u_2}  v^{p_1 q_1}_{r_1 s_1}
    \\
    \label{eq:s12_v2222}
    \langle \sigma_{12} v^{22}_{22} \rangle &=&  
    (-1)^{n_{j_1}}  \rho_{t_1} \, \rho_{p_2 q_2}^{u_2 s_2 r_2} \, \sigma_{t_1 u_2}  v^{p_2 q_2}_{r_2 s_2}
    \\
    \langle \sigma_{12} v^{22}_{11} \rangle &=&  
    (-1)^{n_{j_1}}  \rho_{t_1}^{s_{1} r_1} \, \rho_{p_2 q_2}^{u_2} \, \sigma_{t_1 u_2}  v^{p_2 q_2}_{r_1 s_1}
    \\
    \langle \sigma_{12} v^{11}_{21} \rangle &=&  
    -2\,  \rho_{p_1 q_1 t_1}^{s_1} \, \rho^{u_2 r_2} \, \sigma_{t_1 u_2}  v^{p_1 q_1}_{r_2 s_1}
    \\
    \langle \sigma_{12} v^{12}_{22} \rangle &=&  
    -2\,  \rho_{p_1 t_1} \, \rho_{q_2}^{u_2 s_2 r_2} \, \sigma_{t_1 u_2}  v^{p_1 q_2}_{r_2 s_2}
    \\
    \langle \sigma_{12} v^{11}_{22} \rangle &=&  
    (-1)^{n_{j_1}}  \rho_{p_1 q_1 t_1} \, \rho^{u_2 s_2 r_2} \, \sigma_{t_1 u_2}  v^{p_1 q_1}_{r_2 s_2}
    \\
    \langle \sigma_{1 2} \sigma_{2 1} v^{1 2}_{1 2} \rangle &=&
    -4 \, \rho_{p_1 t_1}^{w_1 r_1} \,\rho_{q_2 v_2}^{u_2 s_2} \,\sigma_{t_1 u_2} \,\sigma_{v_2 w_1} \,v^{p_1 q_2}_{r_1 s_2}
    \\
    \label{eq:s12_s21_v1111}
    \langle \sigma_{1 2} \sigma_{2 1} v^{1 1}_{1 1} \rangle &=&
    -\rho_{p_1 q_1 t_1}^{w_1 s_1 r_1} \,\rho_{v_2}^{u_2} \,\sigma_{t_1 u_2} \,\sigma_{v_2 w_1} \,v^{p_1 q_1}_{r_1 s_1}
    \\
    \langle \sigma_{1 2} \sigma_{1 2} v^{2 2}_{1 1} \rangle &=&
    \frac{1}{2}\rho_{t_1 v_1}^{s_1 r_1} \,\rho_{p_2 q_2}^{w_2 u_2} \,\sigma_{t_1 u_2} \,\sigma_{v_1 w_2} \,v^{p_2 q_2}_{r_1 s_1}
    \\
    \langle \sigma_{1 2} \sigma_{1 2} v^{1 2}_{1 1} \rangle &=&
    \\ &~&
    (-1)^{n_{j_1}} \rho_{p_1 t_1 v_1}^{s_1 r_1} \,\rho_{q_2}^{w_2 u_2} \,\sigma_{t_1 u_2} \,\sigma_{v_1 w_2} \,v^{p_1 q_2}_{r_1 s_1} \nonumber
    \\
    \langle \sigma_{1 2} \sigma_{1 2} v^{2 2}_{2 1} \rangle &=&
    \\ &~&
    (-1)^{n_{j_1}} \rho_{t_1 v_1}^{s_1} \,\rho_{p_2 q_2}^{w_2 u_2 r_2} \,\sigma_{t_1 u_2} \,\sigma_{v_1 w_2} \,v^{p_2 q_2}_{r_2 s_1} \nonumber
    \\
    \langle \sigma_{1 2} \sigma_{2 1} v^{1 1}_{2 1} \rangle &=&
    \\ &~&
    -2(-1)^{n_{j_1}} \rho_{p_1 q_1 t_1}^{w_1 s_1} \,\rho_{v_2}^{u_2 r_2} \,\sigma_{t_1 u_2} \,\sigma_{v_2 w_1} \,v^{p_1 q_1}_{r_2 s_1} \nonumber
    \\
    \langle \sigma_{1 2} \sigma_{2 1} v^{1 2}_{1 1} \rangle &=&
    \\ &~&
    -2(-1)^{n_{j_1}} \rho_{p_1 t_1}^{w_1 s_1 r_1} \,\rho_{q_2 v_2}^{u_2} \,\sigma_{t_1 u_2} \,\sigma_{v_2 w_1} \,v^{p_1 q_2}_{r_1 s_1} \nonumber
    \\
    \langle \sigma_{1 2} \sigma_{1 2} v^{1 2}_{1 2} \rangle &=&
    \\ &~&
    2\rho_{p_1 t_1 v_1}^{r_1} \,\rho_{q_2}^{w_2 u_2 s_2} \,\sigma_{t_1 u_2} \,\sigma_{v_1 w_2} \,v^{p_1 q_2}_{r_1 s_2} \nonumber
    \\
    \label{eq:s12_s12_v1111}
    \langle \sigma_{1 2} \sigma_{1 2} v^{1 1}_{1 1} \rangle &=&
    \frac{1}{2}\rho_{p_1 q_1 t_1 v_1}^{s_1 r_1} \,\rho_{}^{w_2 u_2} \,\sigma_{t_1 u_2} \,\sigma_{v_1 w_2} \,v^{p_1 q_1}_{r_1 s_1}
    \\
    \label{eq:s12_s12_v2222}
    \langle \sigma_{1 2} \sigma_{1 2} v^{2 2}_{2 2} \rangle &=&
    \frac{1}{2}\rho_{t_1 v_1} \,\rho_{p_2 q_2}^{w_2 u_2 s_2 r_2} \,\sigma_{t_1 u_2} \,\sigma_{v_1 w_2} \,v^{p_2 q_2}_{r_2 s_2}
    \\
    \langle \sigma_{1 2} \sigma_{2 1} v^{1 1}_{2 2} \rangle &=&
    \\ &~&
    -\rho_{p_1 q_1 t_1}^{w_1} \,\rho_{v_2}^{u_2 s_2 r_2} \,\sigma_{t_1 u_2} \,\sigma_{v_2 w_1} \,v^{p_1 q_1}_{r_2 s_2} \nonumber
    \\
    \langle \sigma_{1 2} \sigma_{1 2} v^{1 1}_{2 1} \rangle &=&
    \\ &~&
    (-1)^{n_{j_1}} \rho_{p_1 q_1 t_1 v_1}^{s_1} \,\rho_{}^{w_2 u_2 r_2} \,\sigma_{t_1 u_2} \,\sigma_{v_1 w_2} \,v^{p_1 q_1}_{r_2 s_1} \nonumber
    \\
    \langle \sigma_{1 2} \sigma_{1 2} v^{1 2}_{2 2} \rangle &=&
    \\ &~&
    (-1)^{n_{j_1}} \rho_{p_1 t_1 v_1}^{} \,\rho_{q_2}^{w_2 u_2 s_2 r_2} \,\sigma_{t_1 u_2} \,\sigma_{v_1 w_2} \,v^{p_1 q_2}_{r_2 s_2} \nonumber
    \\
    \langle \sigma_{1 2} \sigma_{1 2} v^{1 1}_{2 2} \rangle &=&
    \\ &~&
    \frac{1}{2}\rho_{p_1 q_1 t_1 v_1}^{} \,\rho_{}^{w_2 u_2 s_2 r_2} \,\sigma_{t_1 u_2} \,\sigma_{v_1 w_2} \,v^{p_1 q_1}_{r_2 s_2} \nonumber
\end{eqnarray}

In terms of these diagram definitions, also with fragment permutations as discussed, the equations for the monomer and dimer matrix elements in eq.~\ref{eq:excitonic} for XR proper are
\begin{eqnarray}
    \label{eq:H1monomer}
    \langle\psi^{i_{1}}| \hat{H}_{\{1\}} |\psi_{j_{1}}\rangle
    &=& \langle h^{1}_{1} \rangle + \langle v^{11}_{11} \rangle
    \\
    \label{eq:H2dimer}
    \langle\psi^{i_{1}} \psi^{i_{2}}| \hat{H}_{\{1,2\}} |\psi_{j_{1}} \psi_{j_{2}}\rangle
    &=&
    \langle {}^2 u^{1}_{1} \rangle \delta_2
    + \langle {}^1 u^{2}_{2} \rangle \delta_1
    \\ &~&
    +~\langle h^{1}_{2} \rangle 
    + \langle h^{2}_{1} \rangle
    + \langle v^{12}_{12} \rangle \nonumber
    \\ &~&
    +~\langle v^{11}_{21} \rangle
    + \langle v^{12}_{11} \rangle
    + \langle v^{11}_{22} \rangle \nonumber
    \\ &~&
    +~\langle v^{22}_{12} \rangle
    + \langle v^{21}_{22} \rangle
    + \langle v^{22}_{11} \rangle \nonumber
\end{eqnarray}
which uses $\hat{h} = \hat{t} + {}^{\mathbf{m}}\hat{u}$ with the value of $\mathbf{m}$ inferred from the relevant global-Hamiltonian component $\hat{H}_{\mathbf{m}}$
 from the left-hand side
 (per eqs.~\ref{eq:abinitiofragdecomp1} and \ref{eq:abinitiofragdecomp2}).
For $\hat{H}_{\{1,2\}}$, for example,
$\langle {}^\mathbf{m}u^{1}_{2} \rangle = \langle {}^1 u^{1}_{2} \rangle + \langle {}^2 u^{1}_{2} \rangle$.

We now also give the complete equations needed to build the overlap 
 matrices of eqs.~\ref{eq:Smatrix} and \ref{eq:targetoverlap} for monomers and dimers through through fourth order for $\hat{S}$,
and Hamiltonian matrices in eqs.~\ref{eq:Hmatrix} and \ref{eq:targetH} through
 second order.
Here, the value of $\mathbf{m}$ (for ${}^{\mathbf{m}}\hat{u}$ in $\hat{h}$) will be inferred from the relevant full subsystem Hamiltonian $\hat{\mathcal{H}}_{\mathbf{m}}$,
which is, in turn, inferred from the system on which $\hat{\mathcal{H}}$ acts, as discussed after eqs.~\ref{eq:newexcitonic2a}--\ref{eq:newexcitonic2c}.
We will also need straightforward generalizations of the (permuted) diagrams with the transformed integrals of eqs.~\ref{eq:symmh}--\ref{eq:symmv}.
\begin{eqnarray}
    \label{eq:S0monomer}
    \langle\psi^{i_{1}}| \hat{S}^{[0]} |\psi_{j_{1}}\rangle
    &=&
    \delta_1
    \\
    \label{eq:SXmonomer}
    \langle\psi^{i_{1}}| \hat{S}^{[o>0]} |\psi_{j_{1}}\rangle
    &=&
    0
\end{eqnarray}

\begin{eqnarray}
    \label{eq:S0dimer}
    \langle\psi^{i_{1}} \psi^{i_{2}}| \hat{S}^{[0]} |\psi_{j_{1}} \psi_{j_{2}}\rangle
    &=&
    \delta_1 \delta_2 \\
    \label{eq:S1dimer}
    \langle\psi^{i_{1}} \psi^{i_{2}}| \hat{S}^{[1]} |\psi_{j_{1}} \psi_{j_{2}}\rangle
    &=&
      \langle \sigma_{12} \rangle
    + \langle \sigma_{21} \rangle
    \\
    \label{eq:S2dimer}
    \langle\psi^{i_{1}} \psi^{i_{2}}| \hat{S}^{[2]} |\psi_{j_{1}} \psi_{j_{2}}\rangle
    &=&
    \langle \sigma_{1 2}   \sigma_{2 1} \rangle
    \\ &~&
    +~\langle \sigma_{1 2} \sigma_{1 2} \rangle + \langle \sigma_{2 1} \sigma_{2 1} \rangle
    \nonumber
    \\
    \label{eq:S3dimer}
    \langle\psi^{i_{1}} \psi^{i_{2}}| \hat{S}^{[3]} |\psi_{j_{1}} \psi_{j_{2}}\rangle
    &=&
    ~~~\langle \sigma_{1 2}   \sigma_{1 2}   \sigma_{2 1} \rangle
    \\ &~&
    +~ \langle \sigma_{1 2}   \sigma_{1 2}   \sigma_{1 2} \rangle \nonumber
    \\ &~&
    +~ \langle \sigma_{2 1}   \sigma_{2 1}   \sigma_{1 2} \rangle \nonumber
    \\ &~&
    +~ \langle \sigma_{2 1}   \sigma_{2 1}   \sigma_{2 1} \rangle \nonumber
    \\
    \label{eq:S4dimer}
    \langle\psi^{i_{1}} \psi^{i_{2}}| \hat{S}^{[4]} |\psi_{j_{1}} \psi_{j_{2}}\rangle
    &=&
    ~~~\langle \sigma_{1 2}   \sigma_{1 2}   \sigma_{2 1}   \sigma_{2 1} \rangle
    \\ &~&
    +~ \langle \sigma_{1 2}   \sigma_{1 2}   \sigma_{1 2}   \sigma_{2 1} \rangle \nonumber
    \\ &~&
    +~ \langle \sigma_{1 2}   \sigma_{1 2}   \sigma_{1 2}   \sigma_{1 2} \rangle \nonumber
    \\ &~&
    +~ \langle \sigma_{2 1}   \sigma_{2 1}   \sigma_{2 1}   \sigma_{1 2} \rangle \nonumber
    \\ &~&
    +~ \langle \sigma_{2 1}   \sigma_{2 1}   \sigma_{2 1}   \sigma_{2 1} \rangle \nonumber
\end{eqnarray}

\begin{eqnarray}
    \label{eq:S0HmonomerA}
    \langle\psi^{i_{1}}| \hat{S}^{[0]}\hat{\mathcal{H}} |\psi_{j_{1}}\rangle
    &=&
    \langle h^{1}_{1} \rangle + \langle v^{11}_{11} \rangle
    \\
    \label{eq:S0HmonomerB}
    &=&
    \langle h_{11} \rangle + \langle v_{1111} \rangle
    \\
    \label{eq:S1HmonomerA}
    \langle\psi^{i_{1}}| \hat{S}^{[1]} \hat{\mathcal{H}} |\psi_{j_{1}}\rangle 
    &=&
    \langle h_{11} \rangle
    -   \langle h^{1}_{1} \rangle \nonumber
    \\ &~&
    +~2 \big(
    \langle v^{1}_{111} \rangle     
    - \langle v^{11}_{11} \rangle
    \big)  \\
    \label{eq:S1HmonomerB}
    &=&
    0 \\
    \label{eq:SXHmonomer}
    \langle\psi^{i_{1}}| \hat{S}^{[o>0]} \hat{\mathcal{H}} |\psi_{j_{1}}\rangle 
    &=&
    0
\end{eqnarray}

\begin{eqnarray}
    \label{eq:S0Hdimer}
    \langle\psi^{i_{1}} \psi^{i_{2}}| \hat{S}^{[0]} \hat{\mathcal{H}} |\psi_{j_{1}} \psi_{j_{2}}\rangle
    &=&
    ~~~ \big(
    \langle h^{1}_{1} \rangle
    + \langle v^{11}_{11} \rangle
    \big) \delta_2
    \\ &~&
    +~\big(
    \langle h^{2}_{2} \rangle
    + \langle v^{22}_{22} \rangle
    \big) \delta_1 \nonumber
    \\ &~&
    +~\langle h^{1}_{2} \rangle
    + \langle h^{2}_{1} \rangle
    + \langle v^{12}_{12} \rangle \nonumber
    \\ &~&
    +~\langle v^{11}_{21} \rangle
    + \langle v^{12}_{11} \rangle
    + \langle v^{11}_{22} \rangle \nonumber
    \\ &~&
    +~\langle v^{22}_{12} \rangle
    + \langle v^{21}_{22} \rangle
    + \langle v^{22}_{11} \rangle \nonumber
\end{eqnarray}

\begin{eqnarray}
    \label{eq:S1Hdimer}
    &~& \langle\psi^{i_{1}} \psi^{i_{2}}| \big[ \hat{S}^{[0]} + \hat{S}^{[1]} \big] \hat{\mathcal{H}} |\psi_{j_{1}} \psi_{j_{2}}\rangle
    \\ &=& \nonumber
    \\ &~&
    ~~~\big(
    \langle h_{11} \rangle
    + 2 \langle v^{1}_{111} \rangle
    - \langle v^{11}_{11} \rangle
    \big) \delta_2 \nonumber
    \\ &~&
    +~\big(
    \langle h_{22} \rangle
    + 2 \langle v^{2}_{222} \rangle
    - \langle v^{22}_{22} \rangle
    \big) \delta_1 \nonumber
    \\ &~&
    +~\langle h_{12} \rangle
    + \langle h_{21} \rangle \nonumber
    \\ &~&
    +~ 2\big(
        \langle v^{1}_{212} \rangle \nonumber
    \\ &~&
    ~ + \langle v^{1}_{121} \rangle 
      + \langle v^{1}_{211} \rangle
      + \langle v^{1}_{122} \rangle \nonumber
    \\ &~&
    ~ + \langle v^{2}_{212} \rangle 
      + \langle v^{2}_{122} \rangle
      + \langle v^{2}_{211} \rangle
    \big) \nonumber
    \\ &~&
    -~ \big(
        \langle v^{12}_{12} \rangle \nonumber
    \\ &~&
    ~ + \langle v^{11}_{21} \rangle
      + \langle v^{12}_{11} \rangle
      + \langle v^{11}_{22} \rangle \nonumber
    \\ &~&
    ~ + \langle v^{22}_{12} \rangle
      + \langle v^{21}_{22} \rangle
      + \langle v^{22}_{11} \rangle
    \big) \nonumber
    \\ &~&
    +~\langle \sigma_{12} h^{2}_{1} \rangle
    + \langle \sigma_{12} h^{1}_{1} \rangle 
    + \langle \sigma_{12} h^{2}_{2} \rangle
    + \langle \sigma_{12} h^{1}_{2} \rangle
    \nonumber
    \\ &~&
    +~\langle \sigma_{21} h^{1}_{2} \rangle
    + \langle \sigma_{21} h^{2}_{2} \rangle
    + \langle \sigma_{21} h^{1}_{1} \rangle
    + \langle \sigma_{21} h^{2}_{1} \rangle
    \nonumber
    \\ &~&
    +~\langle \sigma_{12} v^{12}_{11} \rangle
    + \langle \sigma_{12} v^{22}_{21} \rangle
    + \langle \sigma_{12} v^{12}_{12} \rangle \nonumber
    \\ &~&
    +~\langle \sigma_{12} v^{11}_{11} \rangle
    + \langle \sigma_{12} v^{22}_{22} \rangle
    + \langle \sigma_{12} v^{22}_{11} \rangle \nonumber
    \\ &~&
    +~\langle \sigma_{12} v^{11}_{21} \rangle
    + \langle \sigma_{12} v^{12}_{22} \rangle
    + \langle \sigma_{12} v^{11}_{22} \rangle \nonumber
    \\ &~&
    +~\langle \sigma_{21} v^{21}_{22} \rangle
    + \langle \sigma_{21} v^{11}_{12} \rangle
    + \langle \sigma_{21} v^{21}_{21} \rangle \nonumber
    \\ &~&
    +~\langle \sigma_{21} v^{22}_{22} \rangle
    + \langle \sigma_{21} v^{11}_{11} \rangle
    + \langle \sigma_{21} v^{11}_{22} \rangle \nonumber
    \\ &~&
    +~\langle \sigma_{21} v^{22}_{12} \rangle
    + \langle \sigma_{21} v^{21}_{11} \rangle
    + \langle \sigma_{21} v^{22}_{11} \rangle \nonumber
\end{eqnarray}

\begin{eqnarray}
    \label{eq:S2Hdimer}
    &~& \langle\psi^{i_{1}} \psi^{i_{2}}| \big[ \hat{S}^{[0]} + \hat{S}^{[1]} + \hat{S}^{[2]} \big] \hat{\mathcal{H}} |\psi_{j_{1}} \psi_{j_{2}}\rangle
    \\ &=& \nonumber
    \\ &~&
    ~~~\big(
    \langle h_{11} \rangle
    + \langle v_{1111} \rangle
    \big) \delta_2
    + \big(
    \langle h_{22} \rangle
    + \langle v_{2222} \rangle
    \big) \delta_1 \nonumber
    \\ &~&
    +~\langle h_{12} \rangle
    + \langle h_{21} \rangle
    + \langle v_{1212} \rangle \nonumber
    \\ &~&
    +~\langle v_{1121} \rangle 
    + \langle v_{1211} \rangle
    + \langle v_{1122} \rangle \nonumber
    \\ &~&
    +~\langle v_{2212} \rangle 
    + \langle v_{2122} \rangle
    + \langle v_{2211} \rangle \nonumber
    \\ &~&
    +~\langle \sigma_{12} h_{21} \rangle
    + \langle \sigma_{12} h_{11} \rangle
    + \langle \sigma_{12} h_{22} \rangle
    + \langle \sigma_{12} h_{12} \rangle \nonumber
    \\ &~&
    +~\langle \sigma_{21} h_{12} \rangle
    + \langle \sigma_{21} h_{22} \rangle
    + \langle \sigma_{21} h_{11} \rangle
    + \langle \sigma_{21} h_{21} \rangle \nonumber
    \\ &~&
    +~ 2\big(
        \langle \sigma_{12} v^{1}_{211} \rangle
      + \langle \sigma_{12} v^{2}_{221} \rangle
      + \langle \sigma_{12} v^{1}_{212} \rangle \nonumber
    \\ &~&
    ~ + \langle \sigma_{12} v^{1}_{111} \rangle
      + \langle \sigma_{12} v^{2}_{222} \rangle
      + \langle \sigma_{12} v^{2}_{211} \rangle \nonumber
    \\ &~&
    ~ + \langle \sigma_{12} v^{1}_{121} \rangle
      + \langle \sigma_{12} v^{1}_{222} \rangle
      + \langle \sigma_{12} v^{1}_{122} \rangle \nonumber
    \\ &~&
    ~ + \langle \sigma_{21} v^{2}_{122} \rangle
      + \langle \sigma_{21} v^{1}_{112} \rangle
      + \langle \sigma_{21} v^{2}_{121} \rangle \nonumber
    \\ &~&
    ~ + \langle \sigma_{21} v^{2}_{222} \rangle
      + \langle \sigma_{21} v^{1}_{111} \rangle
      + \langle \sigma_{21} v^{1}_{122} \rangle \nonumber
    \\ &~&
    ~ + \langle \sigma_{21} v^{2}_{212} \rangle
      + \langle \sigma_{21} v^{2}_{111} \rangle
      + \langle \sigma_{21} v^{2}_{211} \rangle
    \big) \nonumber
    \\ &~&
    -~ \big(
        \langle \sigma_{12} v^{12}_{11} \rangle
      + \langle \sigma_{12} v^{22}_{21} \rangle
      + \langle \sigma_{12} v^{12}_{12} \rangle \nonumber
    \\ &~&
    ~ + \langle \sigma_{12} v^{11}_{11} \rangle
      + \langle \sigma_{12} v^{22}_{22} \rangle
      + \langle \sigma_{12} v^{22}_{11} \rangle \nonumber
    \\ &~&
    ~ + \langle \sigma_{12} v^{11}_{21} \rangle
      + \langle \sigma_{12} v^{12}_{22} \rangle
      + \langle \sigma_{12} v^{11}_{22} \rangle \nonumber
    \\ &~&
    ~ + \langle \sigma_{21} v^{21}_{22} \rangle
      + \langle \sigma_{21} v^{11}_{12} \rangle
      + \langle \sigma_{21} v^{21}_{21} \rangle \nonumber
    \\ &~&
    ~ + \langle \sigma_{21} v^{22}_{22} \rangle
      + \langle \sigma_{21} v^{11}_{11} \rangle
      + \langle \sigma_{21} v^{11}_{22} \rangle \nonumber
    \\ &~&
    ~ + \langle \sigma_{21} v^{22}_{12} \rangle
      + \langle \sigma_{21} v^{21}_{11} \rangle
      + \langle \sigma_{21} v^{22}_{11} \rangle
    \big) \nonumber
    \\ &~&
    +~\langle \sigma_{12} \sigma_{21} h^{1}_{1} \rangle
    + \langle \sigma_{12} \sigma_{21} h^{1}_{2} \rangle
    + \langle \sigma_{12} \sigma_{12} h^{2}_{1} \rangle \nonumber
    \\ &~&
    +~\langle \sigma_{12} \sigma_{12} h^{1}_{1} \rangle
    + \langle \sigma_{12} \sigma_{12} h^{2}_{2} \rangle
    + \langle \sigma_{12} \sigma_{12} h^{1}_{2} \rangle \nonumber
    \\ &~&
    +~\langle \sigma_{21} \sigma_{12} h^{2}_{2} \rangle
    + \langle \sigma_{21} \sigma_{12} h^{2}_{1} \rangle
    + \langle \sigma_{21} \sigma_{21} h^{1}_{2} \rangle \nonumber
    \\ &~&
    +~\langle \sigma_{21} \sigma_{21} h^{2}_{2} \rangle
    + \langle \sigma_{21} \sigma_{21} h^{1}_{1} \rangle
    + \langle \sigma_{21} \sigma_{21} h^{2}_{1} \rangle \nonumber
    \\ &~&
    +~\langle \sigma_{12} \sigma_{21} v^{12}_{12} \rangle
    + \langle \sigma_{12} \sigma_{21} v^{11}_{11} \rangle
    + \langle \sigma_{12} \sigma_{12} v^{22}_{11} \rangle \nonumber
    \\ &~&
    +~\langle \sigma_{12} \sigma_{12} v^{12}_{11} \rangle
    + \langle \sigma_{12} \sigma_{12} v^{22}_{21} \rangle
    + \langle \sigma_{12} \sigma_{21} v^{11}_{21} \rangle \nonumber
    \\ &~&
    +~\langle \sigma_{12} \sigma_{21} v^{12}_{11} \rangle
    + \langle \sigma_{12} \sigma_{12} v^{12}_{12} \rangle
    + \langle \sigma_{12} \sigma_{12} v^{11}_{11} \rangle \nonumber
    \\ &~&
    +~\langle \sigma_{12} \sigma_{12} v^{22}_{22} \rangle
    + \langle \sigma_{12} \sigma_{21} v^{11}_{22} \rangle
    + \langle \sigma_{12} \sigma_{12} v^{11}_{21} \rangle \nonumber
    \\ &~&
    +~\langle \sigma_{12} \sigma_{12} v^{12}_{22} \rangle
    + \langle \sigma_{12} \sigma_{12} v^{11}_{22} \rangle
    + \langle \sigma_{21} \sigma_{12} v^{22}_{22} \rangle \nonumber
    \\ &~&
    +~\langle \sigma_{21} \sigma_{21} v^{11}_{22} \rangle
    + \langle \sigma_{21} \sigma_{21} v^{21}_{22} \rangle
    + \langle \sigma_{21} \sigma_{21} v^{11}_{12} \rangle \nonumber
    \\ &~&
    +~\langle \sigma_{21} \sigma_{12} v^{22}_{12} \rangle
    + \langle \sigma_{21} \sigma_{12} v^{21}_{22} \rangle
    + \langle \sigma_{21} \sigma_{21} v^{21}_{21} \rangle \nonumber
    \\ &~&
    +~\langle \sigma_{21} \sigma_{21} v^{22}_{22} \rangle
    + \langle \sigma_{21} \sigma_{21} v^{11}_{11} \rangle
    + \langle \sigma_{21} \sigma_{12} v^{22}_{11} \rangle \nonumber
    \\ &~&
    +~\langle \sigma_{21} \sigma_{21} v^{22}_{12} \rangle
    + \langle \sigma_{21} \sigma_{21} v^{21}_{11} \rangle
    + \langle \sigma_{21} \sigma_{21} v^{22}_{11} \rangle \nonumber
\end{eqnarray}

We have included
eqs.~\ref{eq:SXmonomer} \& \ref{eq:S1HmonomerA}--\ref{eq:SXHmonomer}
in order
to display complete patterns to aide in following the derivations, and to demonstrate some internal consistencies.
As discussed immediately after
 eqs.~\ref{eq:newexcitonic2a}--\ref{eq:newexcitonic2c},
 the bra states here are biorthogonal complements for the relevant subsystem only.
Combined with the fact that we have assumed an orthonormal basis for the fragments (\textit{i.e.}, a Hermitian $\mathbf{z}$ with orthonormal orbitals within a fragment), it is clear at the highest level of abstraction that $\hat{S}=\hat{S}^{[0]}=1$ for monomer systems, so all higher-order terms for monomers should vanish.
At a more detailed level, eq.~\ref{eq:SXmonomer} yields zero
for any nonzero order,
because diagonal blocks of $\boldsymbol{\sigma}$ are then zero.
Similarly, there is no distinction between the original orbitals and their biorthogonal complements for monomer systems when using XR[$o$] variants under these conditions, causing \ref{eq:S1HmonomerA} to vanish identically, and likewise for any higher-order terms.
To emphasize this latter point, we also write the integrals
 in the symmetric representation for the XR[0] monomer Hamiltonian elements in eq.~\ref{eq:S0HmonomerB}.
It is not strictly necessary to do this, since the equivalence is implicit for XR[0], but it also emphasizes the difference in monomer Hamiltonian relative to XR
in eq.~\ref{eq:H1monomer}, wherein the bra orbitals for the integrals are orthogonal to the original orbitals on all other fragments.
The difference between XR and XR[0]
 for complete dimer Hamiltonians (including also monomer terms for XR)
 is left implicit,
 as our notation does not distinguish whether bra orbitals are biorthogonal complements to only the subsystem orbitals or to the global set of supersystem orbitals.

\section{\label{sec:nonhermitianfrags} Further Generalization of the XR Model}

All variants of excitonic renormalization are simply dependent on a choice of complements that can be straightforwardly applied in eq.~\ref{eq:newexcitonic1} (via eqs.~\ref{eq:newexcitonic2a}--\ref{eq:newexcitonic2c}, \textit{etc.}).
If we had a set of states, like EOM-CC states, for which we (implicitly) know the inverse of $\mathbf{z}$ and $\mathbf{Z}$ but for which we cannot take proper expectation values, we can define a set of states according to
\begin{eqnarray}
    \langle\Psi'_I| = \sum_P \bar{Z}^I_P \langle\Phi_P|
\end{eqnarray}
which can be thought of as ``partial'' complements, with respect to coefficients but not orbitals.
As with $\bar{\mathbf{S}}$, we use the overbar to denote an inverse and its elements, \textit{e.g.}, $\bar{Z}^I_P \in \bar{\mathbf{Z}}=\mathbf{Z}^{-1}$.
This has the property that $|\Psi'_I\rangle = |\Psi_I\rangle$ if $\mathbf{z}$ (and therefore $\mathbf{Z}$) is Hermitian, and this state can replace $|\Psi_I\rangle$ as the bra in the definition of $\mathbf{S}$ in eq.~\ref{eq:Smatrix} and in the right-hand side of eq.~\ref{eq:newcomplements}.
Note that the new definition of $\mathbf{S}$ changes its interpretation, no longer being a Hermitian matrix of overlaps between equivalent bra and ket states.
Nevertheless the new definition of $\{|\bar{\Psi}^{\tilde{I}}\rangle\}$ from eq.~\ref{eq:newcomplements} is immediately seen to be a more general set of complements, which are the same as the ones we have been discussing if $\mathbf{z}$ and $\mathbf{Z}$ are Hermitian.

With this definition, one can see in eq.~\ref{eq:newexcitonic2b} that the monomer Hamiltonians according to the original level of theory for the fragments are thus reproduced and that $|\Psi'_I\rangle$ should also replace $|\Psi_I\rangle$ as the bra in the definition of $\tilde{\mathbf{H}}$ in eq.~\ref{eq:Hmatrix}.
Furthermore these ``prime'' analogues of the monomer states should appear in the bra of eq.~\ref{eq:covarianttenspdt}
 and also in the left-hand sides of eqs.~\ref{eq:targetoverlap} and \ref{eq:targetH}, but the right-hand sides of the latter two equations do not change; we have simply lifted the restriction that $\mathbf{z}$ and $\mathbf{Z}$ are Hermitian (by bringing the left-hand side into alignment with the more general framework).

Finally, the elements of $\mathbf{z}$ on the right-hand side of eq.~\ref{eq:fieldopsandwich} should be replaced with unconjugated elements of $\bar{\mathbf{z}}$ with the indices transposed (again, this represents no change if $\mathbf{z}$ and $\mathbf{Z}$ are Hermitian).
Also, the $i_m$ in eq.~\ref{eq:transdensities} should remain as a superscript to remind us that, although we need not contend with a biorthogonal orbital basis, we should use the biorthogonal complements (at the coefficient level) from the original fragment theory.

\end{document}